\documentclass[a4paper,fleqn,usenatbib]{mnras}

\usepackage{multicol}
\usepackage[T1]{fontenc}
\usepackage{ae,aecompl}

\usepackage{graphicx}	
\usepackage{amsmath}	
\usepackage{amssymb}	

\title[TESS light curves of cataclysmic variables -- I]
{TESS light curves of cataclysmic variables -- I -- Unknown periods in 
long-known stars}

\author[A. Bruch]{Albert Bruch
\\
Laborat\'orio Nacional de Astrof\'{\i}sica, Rua Estados Unidos, 154, 
CEP 37500-364, Itajub\'a, MG, Brazil
}

\date{Accepted XXX. Received YYY; in original form ZZZ}

\pubyear{2019}

\newcommand{\hochpunkt}[1]{\mbox{$^{\raisebox{.3ex}{\scriptsize #1}}_{\raisebox{.6ex}{\hspace{.17em}.}}$}}

\begin{document}
\label{firstpage}
\pagerange{\pageref{firstpage}--\pageref{lastpage}}
\maketitle

\begin{abstract}
Cataclysmic variables (CVs) 
exhibit a plethora of variable phenomena many of which
require long, uninterrupted light curves to reveal themselves in detail. The
month long datasets provided by TESS are well suited for this purpose. TESS
has the additional advantage to have observed a huge number of stars, among
them many CVs. Here, a search for periodic variations in a sample of CVs
of the novalike and old novae subtypes is presented. In 10
of the 15 targets either previously unseen positive or negative superhumps
or unusual features in known superhumps are identified. The TESS light curves 
demonstrate that the occurrence of superhumps in these types of CVs is not an
exception but quite common. For 8 systems new or
improved values for the orbital period are measured. In TV~Col the long-sought
optical manifestation of the white dwarf spin period is first seen in form of
its orbital sideband. The mystery of multiple photometric periods observed
in CP Pup in the past is explained by irregularly occurring anomalous states
which are reflected in the light curve. 
\end{abstract}




\section{Introduction}
\label{Introduction}

Cataclysmic variables (CVs) are well known as interacting binary stars
where a Roche-lobe filling late type (secondary) star which is in most 
cases on or close to the main sequence orbits around a more massive 
(primary) white dwarf (WD) with a period ranging between $\approx$75~min and 
rarely more than 10~h. Matter is transferred from the secondary to the
primary either via an accretion disk, through magnetic interaction, or 
a combination of both. For a comprehensive discussion of most aspects of
CVs, see \citet{Warner95}. 

The dynamics of the various stellar and non-stellar (gaseous) components of 
CVs give rise to a plethora of phenomena which lead to the emission of 
radiation over the entire electromagnetic spectrum (radio -- $\gamma$-rays)
and which is strongly variable on all time scales between seconds and 
millennia. 
Concentrating on optical light, the variability may present itself as
stochastic (e.g., flickering on time scales of minutes), irregular (e.g.,
high- and low states), semi-regular (e.g., quasi-periodic oscillations), 
episodic (e.g., dwarf nova or classical nova outbursts), or periodic
(e.g., orbital variations). I am not aware of any other class of astronomical
objects which exhibits a richer phenomenology of variability than CVs. 

Even if some of the variations appear to be periodic, strictly speaking, none
really is, not even orbital variations or modulations due to the spin of the
WD. Here, I consider a variation as periodic
if it is detectable with a stable period over the course of a single
observing mission of a given star. In this sense, variations due to the
orbital motion and the WD rotation are, of course, always periodic.
Other periodic phenomena (albeit less stable than orbit and spin)
often observed in CVs are the so called superhumps (SHs)
which come in two flavours, as positive superhump (pSH) and negative
superhump (nSH), depending on whether the difference between the superhump
and the orbital period is positive or negative.

Positive superhumps
are routinely observed in superoutbursts of SU~UMa type dwarf
novae. They are explained by tidal stresses in an accretion disk which
has become elliptical as a result of a 3:1 resonance between the rotational
period of matter in the outer accretion disk and the orbital period of
the secondary star in a CV \citep{Whitehurst88, Whitehurst91}. 
The period $P_{\rm pSH}$ is slightly
longer than the orbital period $P_{\rm orb}$ because of apsidal precession
of the elliptic disk. The superhump, orbital and precession periods are
related to each other by $1/P_{\rm prec, abs} = 1/P_{\rm orb} - 1/P_{\rm pSH}$.
Although most common in short period dwarf novae in superoutburst, pSHs
are also seen in an increasing number of longer period CVs, independent of
outbursts; mostly in old novae or novalike variables where they can persist 
over long periods of time or occur only intermittently. Theory predicts that the
disk can reach the 3:1 resonance only in systems with a small mass ratio
$q=M_{\rm sec}/M_{\rm prim}$. Just how small $q$ must be is a matter of 
debate. \citet{Whitehurst91} estimate a maximum of 
$q_{\rm max} = 0.33$\footnote{\citet{Pearson06} points out an error in
eq.~5 of \citet{Whitehurst91} and correct this value to $q_{\rm max}=0.28$.}.
a values often quoted in the literature. \citet{Pearson06}, on the other
hand, gets $q_{\rm max}=0.39$, while \citet{Smak20} advocates a small value
of $q_{\rm max}=0.22$. Since, on average, $q$ increases with the orbital 
period the condition for pSHs to arise will be more difficult to achieve
for long period systems. Assuming an average white dwarf mass of 
0.83~$M_\odot$ \citep{Zorotovic11} and a secondary star mass according to
the semi-empirical CV donor sequence of \citet{Knigge11} pSHs are then
limited to systems with an orbital period below 2.1~h for $q_{\rm max}=0.22$
[which causes \citet{Smak20} to argue that the canonical model fails to 
explain pSHs in CVs above the period gap] and 4.1~h for $q_{\rm max}=0.39$.

In contrast, negative superhumps occur in systems with a 
warped accretion disk \citep{Thomas15}. 
Depending on the aspect between the disk and the
infalling stream of matter from the secondary the latter penetrates more or 
less deeply into the gravitational well of the WD before hitting the disk, 
modulating thus the release of energy. The warped disk suffers a retrograde 
precession of the nodal line because of tidal torques \citep{Montgomery09}
which means that the released energy and thus
the brightness is modulated with a period slightly smaller than the orbital
period. Again, we have the relation between the superhump, orbital and
precession period $1/P_{\rm prec, nod} = 1/P_{\rm nSH} - 1/P_{\rm orb}$. There is
no consensus about the mechanism which leads to warped accretion disks
\citep{Thomas15}. Therefore, it cannot be
predicted under which circumstances nSHs should or should not develop.

The detection of SHs (or any other periodic signal) in the light curves of 
CVs depends basically on two
parameters: 1) their amplitude with respect to competing sources of 
variability (flickering, data noise), and (2) the temporal properties of
the light curve (time resolution, total time base, gaps). 
The larger the amplitude, and the
longer and free of gaps the light curve, the easier is their detection.
Terrestrial observations suffer from the limited length of nightly light
curves. They cover at most 2 or 3
cycles of periods which are often typically of the order of hours,
and it may be difficult to distinguish between different sources of
variability. If observations from several nights are stitched together the
light curves contain long gaps which lead to strong aliasing patterns of
sometimes difficult interpretation in the power spectra which are generally 
used to find periods. Such complications are avoided using the month-long 
(almost) gap-free light curves generated by the Transiting Exoplanet
Survey Satellite (TESS) \citep{Ricker14} which have
become available in recent years for a great number of CVs.

Here, I take advantage of TESS light curves of a sample of novalike 
variables (NLs) and old novae in order to search for previously unknown periodic
signals in their light curves, to confirm published reports of periods and
improve their numerical values, and to document their evolution and changes 
with respect to earlier epochs. Based on a quick look at the light 
curves and power spectra of many systems, 15 targets were selected for the
present study because it appeared to be particularly promising that a closer
look would reveal interesting features. Of these, five are known to exhibit
superhumps. As it will be shown, another five targets are newly identified
as superhumpers. In a subsequent paper (Bruch, in preparation) the available
TESS data of all remaining NLs and old novae with reports in the literature 
to show superhumps will be investigated in order to gain a better 
understanding of the behaviour of the entire ensemble of such systems.

\section{Data and data handling}
\label{Data and data handling}

The vast majority of data used in this study was observed with the TESS
satellite and downloaded from the Barbara A. Misulski Archive for Space
Telescopes (MAST)\footnote{https://archive.stsci.edu}. For two of the
target objects, data from the Kepler and the subsequent K2 missions are
also available and were downloaded from the same source. For one object
the TESS data were supplemented by data obtained from the American
Association of Variable Star Observers (AAVSO) 
arquives\footnote{https://www.aavso.org}.

TESS observes different sectors of the sky continuously for two spacecraft
orbits of about 27 days. Depending on their location on the sky different 
sectors may overlap each other. 
This means that a given object may be included in more than one
sector. If these sectors are observed in immediate succession, the light
curves of these objects may be combined and then extend over a longer time 
interval. Moreover,
some of the sectors have been observed more than once, generating light
curves at different epochs. Here, I designate light curves of a given star, 
derived from observations of one or more sectors adjacent in time at a given
epoch as LC\#1, LC\#2, etc. The time resolution is 2~min throughout. 
All light curves contain gaps of a few days
used for data transfer to Earth or, in some cases, technical problems.
The start and end epochs of the light curves of individual objects are 
listed in 
Table~1. When comparing TESS data to observations taken with other instruments
it should be kept in mind that its passband encompasses a
wide range between 6\,000 and 10\,000~\AA, centred on the Cousins $I$-band.
Kepler has a similarly broad passband, but offset by roughly 1\,000~\AA\ to
the blue.

MAST offers the TESS (and Kepler) data in different forms. Light curves
are provided as Simple Aperture Photometry (SAP) or Pre-Search Data
Conditioning Simple Aperture Photometry (PDCSAP). A critical comparison
is provided by \citet{Kinemuchi12} for Kepler data. But the
situation for TESS is not very different. PDCSAP data result from
a pipeline which attempt to mitigate instrumental effects. However, in
certain cases this can do more harm than good. In particular I find that
light curves of dwarf novae containing outbursts (not used here) can be
completely distorted by the conditioning process. Moreover, as
\citet{Kinemuchi12} point out, the instrumental effects do not affect
the frequency range $>$1~d$^{-1}$ in which I am most interested here, 
and data analysis of CVs can successfully be done using SAP data. 
Therefore, I do not consider PDCSAP data. 

The main purpose of this study is the search of periodic variations in
the target stars. Therefore, I make ample use of Fourier techniques to
calculate periodograms (hereafter also termed power spectra) applying
the Lomb-Scargle algorithm \citep{Lomb76, Scargle82} or following
\citet{Deeming75}. Both yield equivalent results. In order for variations
on longer time scales not to interfere these were, in general, first
removed by subtraction of a \citet{Savitzky64} filtered version of the
light curve, using a cut-off time scale of 2~d and a 4$^{\rm th}$ order
smoothing polynomial. This was, of course, not done when variations on
longer time scales were considered. In systems with deep eclipses these 
are masked before calculating power spectra.

Througout this paper I use the notation $F_{\rm orb}$, and $F_{\rm SH}$ for
orbital and superhump frequencies ($F_{\rm pSH}$ for positive and $F_{\rm nSH}$
for negative superhumps if they need to be distinguished). The white dwarf 
rotation frequency is denoted by $F_{\rm spin}$. $P$ is used for the 
corresponding periods. Most of the periods investigated here are derived from
the frequency of peaks in power spectra. Their errors are estimated according
to the recipe in Sect.~4.4 of \citet{Schwarzenberg-Czerny91}. This relies
on the pseudo-continuum close to the spectral peak to be caused by data
noise. However, flickering and window patterns of real oscillations may
increase the level of the pseudo-continuum, leading to an overestimation of
the error. Thus, all errors based on this technique can be regarded as
conservative upper limits.
To simplify notation, errors are always given in brackets in units of the
last decimal digits of the nominal value of the frequency/period.

\begin{table}
\label{Table: obs-log}	\centering
	\caption{Journal of observations.}

\begin{tabular}{lccc}
\hline

Name & LC     & Start time & End time \\
     & number & \multicolumn{2}{c}{BJD 2450000+} \\
\hline
V704 And  & 1 & 8764.69 & 8789.68 \\ [1ex]
TT Ari    & 1 & 9447.69 & 9498.88 \\ [1ex]
AC Cnc    & 1$^a$ & 7139.60 & 7214.44 \\
          & 2$^a$ & 8251.54 & 8302.40 \\
          & 3 & 9500.20 & 9550.63 \\ [1ex]
QU Car    & 1 & 9307.26 & 9350.55 \\ [1ex]
V504 Cen  & 1 & 9333.87 & 9360.54 \\ [1ex]
TV Col    & 1 & 8437.99 & 8490.05 \\
          & 2 & 9174.23 & 9217.57 \\ [1ex]
DM Gem    & 1 & 9474.17 & 9576.53 \\ [1ex]
RZ Gru    & 1 & 9061.86 & 9087.26 \\ [1ex]
V533 Her  & 1 & 9010.27 & 9035.14 \\
          & 2 & 9310.66 & 9418.86 \\ [1ex]
V795 Her  & 1 & 8983.64 & 9035.14 \\ [1ex]
HR Lyr    & 1 & 9390.66 & 9418.86 \\ [1ex]
MV Lyr    & 1$^b$ & 5002.51 & 6424.01 \\
          & 2 & 8683.35 & 8710.21 \\
          & 3 & 9010.27 & 9025.14 \\
          & 4 & 9390.66 & 9446.48 \\ [1ex]
AQ Men    & 1 & 8325.30 & 8353.18 \\
          & 2 & 8437.99 & 8464.26 \\
          & 3 & 8517.36 & 8542.00 \\
          & 4 & 8596.78 & 8682.86 \\
          & 5 & 9036.28 & 9092.89 \\
          & 6 & 9144.51 & 9169.94 \\
          & 7 & 9254.11 & 9279.98 \\
          & 8 & 9333.80 & 9389.72 \\ [1ex]
V1193 Ori & 1 & 8437.99 & 8464.00 \\
          & 2 & 9174.23 & 9200.23 \\ [1ex]
CP Pup    & 1 & 8492.22 & 8542.00 \\
          & 2 & 9228.77 & 9279.98 \\ [1ex]
\hline
\multicolumn{4}{l}{$^a$Kepler K2 data} \\
\multicolumn{4}{l}{$^b$Kepler data}
\end{tabular}
\end{table}

\section{Results}

\subsection{V704 And: The correction of the orbital period and a strong 
negative superhump}

V704~And (= LD~317) was discovered as a blue variable star by 
\citet{Dahlmark99}. The first more detailed observations were published by
\citet{Papadaki06} who, based on different brightness levels seen
at different epochs, classified the star as a novalike variable
of the VY~Scl type. This was confirmed by \citet{Weil18}
spectroscopically and through an analysis of the long term light curve.
These authors also measured the orbital period. The time resolved light curves
of \citet{Papadaki06} did not yield convincing evidence of variations
on the orbital or similar periods. This is different in the present
data.

\begin{figure}
	\includegraphics[width=\columnwidth]{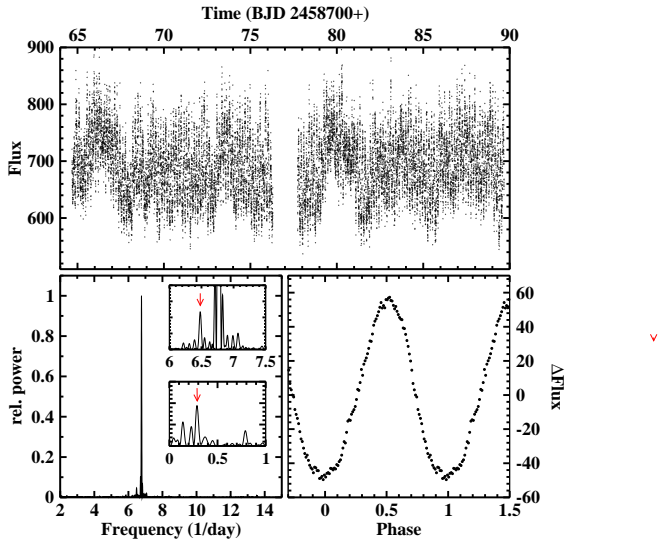}
      \caption[]{{\it Top:} Light curve of V704~And.
                {\it Bottom left:} Power spectrum of the light curve. 
                The upper insert 
                shows the region around the main peak on an expanded scale,
                with the orbital signal marked by an arrow. The lower insert
                contains the low frequency range ($>$2~d variations not 
                removed). The arrow marks the beat between the strongly 
                dominating signal at 6.7696~d$^{-1}$
                and the orbital frequency. 
                {\it Bottom right:} Light curve folded on the period
                0.147\,72~d,
                binned in phase intervals of width 0.01.}
\label{v704and-lc}
\end{figure}

V704~And was in the high state when TESS looked at it, as evidenced
by AAVSO long term observations. The light curve is plotted in the upper
frame of Fig.~\ref{v704and-lc}. It is characterized by strong short term
modulations and variations on time scales of days  which appear irregular
to the eye.

The power spectrum
(lower left frame of Fig.~\ref{v704and-lc}) is dominated by a very strong
signal at 
6.7696~(8)~d$^{-1}$.The first overtone is also seen, but is 
quite faint. The corresponding period of 
$P_1 = 0.147\,72\, (2) $~d is shorter than the spectroscopic orbital period
$P_{\rm WTH} = 0.151\,424\,(3)$~d measured by \citet{Weil18}. The only other 
possibly significant signal (other than side lobes of the main one) is a 
much smaller peak at 
$F_2 = 6.481\,(4)\, {\rm d}^{-1}$ ($P_2 = 0.1542\,(1)$~d) 
(marked in the upper
insert in the figure). This is closer to, but 
still markedly different from $P_{\rm WTH}$. The light curve, folded on $P_1$
and binned in phase intervals of width 0.01
is shown in the right lower frame of Fig.~\ref{v704and-lc}. The waveform can
well be approximated by a simple sine curve.


There is reason to suspect a misprint in period $P_{\rm WTH}$ cited by
\citet{Weil18}. Assuming that the third decimal digit in
$P_{\rm WTH} = 0.151\,424$~d erroneously slipped in, we are left with
$P_{\rm WTH,corr} = 0.154\,24$ which is identical to $P_2$ within the error
of the latter. This by itself is, of course, not sufficient evidence for
an error. However, inspecting fig.~4 of \citet{Weil18} it is 
obvious that the dominant peak in their periodogram is just below 
6.5~d$^{-1}$, compatible with $1/P_2$, but not with 
$1/P_{\rm WTH}=6.603\,97$~d$^{-1}$. This lends more credibility to a misprint.
Noting that the period of \citet{Weil18} was derived from a longer time base
than $P_2$ and should thus be more precise, I will therefore henceforth adopt 
$P_2 \approx P_{\rm WTH,corr} = P_{\rm orb} = 0.154\,24\,(3)\,{\rm d}$ 
as the orbital period of V704~And.

The dominant photometric period $P_1$, coherent over the entire time base 
of the TESS observations, is thus distinctly different from the orbital
period. These photometric variations therefore constitute a (negative, since 
$P_1 \equiv P_{\rm SH} < P_{\rm orb}$) superhump. 
%

\begin{figure}
	\includegraphics[width=\columnwidth]{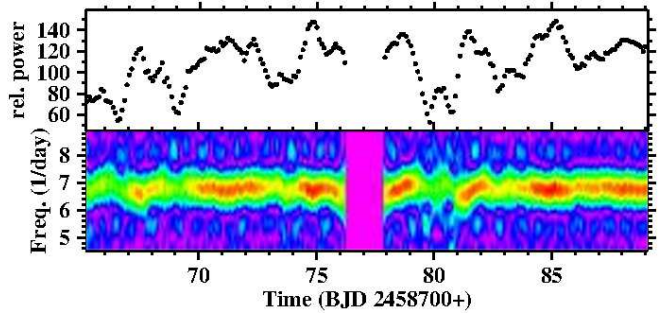}
      \caption[]{{\it Bottom:} Time resolved power spectrum of a narrow
                frequency range around the superhump frequency in V704~And.
                {\it Top:} Power integrated over the width of the superhump
                signal in the lower frame.}
\label{v704and-stacked}
\end{figure}

Further investigation of the superhump signal confirms this interpretation.
The lower frame of Fig.~\ref{v704and-stacked} shows the time resolved power 
spectrum in the frequency range around $1/P_{\rm SH}$, constructed according
to \citet{Bruch14}, and using a sliding window of width 1~d.
The power of the superhump signal is clearly modulated with time as
quantified by the graph in the upper frame of the figure. The power spectrum
of the latter contains a strong peak corresponding to 
3.46~(5)~d. 
Moreover, 
although not immediately apparent, the original data are
also modulated at the same period, as evidenced by the low frequency part
of their power spectrum (lower insert in the lower left frame of 
Fig.~\ref{v704and-lc} where the corresponding frequency is marked by an 
arrow). The period is
3.48 (4)~d which is fully compatible with the beat period of 
3.49~(2)~d between orbital and superhump periods. 

\subsection{TT Ari: Negative and positive superhumps simultaneously present}

TT~Ari, detected as a variable star by \citet{Strohmeier57}, is undoubtedly 
one of the best studied novalike variables. Its 
photometric history over the past 40~y has recently been scrutinized by
\citet{Bruch19} who also provides ample references to previous work. It
may therefore be surprising that TESS data can add yet new aspects
to the rich photometric behaviour of this system. As a VY~Scl type star
TT~Ari alternates between more common high states and occasional states
of reduced brightness. The spectroscopically determined orbital period
of 0.137\,550\,40~(17)~d \citep{Wu02} causes a photometric modulation only
during the deep low state \citep{Bruch19}. In the high state it has never
been detected. Instead, the optical light is dominated by a permanent 
negative superhump with an average period of 0.132\,95~d which slightly
varies over time. This behaviour was, however, interrupted between 1997
and 2004, when the negative was replaced by a positive superhump with
an average period of 0.149\,07~d \citep[see][for details]{Bruch19}.

\begin{figure}
	\includegraphics[width=\columnwidth]{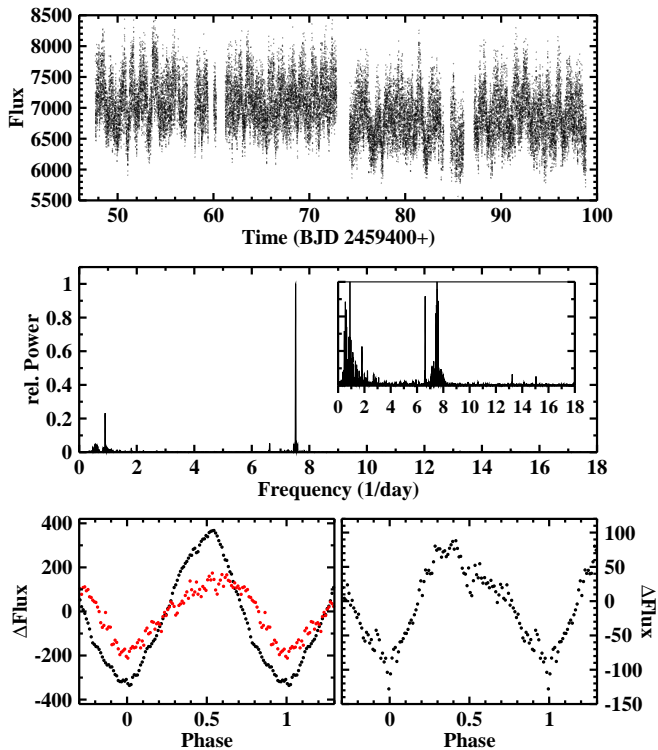}
      \caption[]{{\it Top:} Light curve of TT~Ari. {\it Middle:} Power spectrum
                of the light curve. The insert shows the same data on an
                expandend vertical scale. {\it Bottom left:} Light curve
                of TT~Ari folded on the period of the negative superhump
                (black) and the beat period between negative and positive
                superhump (red), binned in phase intervals of width 0.01.
                {\it Bottom right:} The same, folded on the period of the
                positive superhump.} 
\label{ttari}
\end{figure}

The TESS light curve of TT~Ari, based on observations in two sectors in 
subsequent time intervals, is shown in the
upper frame of Fig.~\ref{ttari}. Apart from slight variations on longer
time scales there appears to be a modulation with
a period of just over 1~d. Therefore, in order to fully maintain these
variations, for the subsequent frequency analysis
the cutoff time scale of the Savitzky-Golay filter used to remove longer
term variations (see Sect.~\ref{Data and data handling}) was increased 
from the usual 2~d to 5~d. The resulting
power spectrum is reproduced in the middle frame of Fig.~\ref{ttari} where the
insert show the same spectrum on an expanded vertical scale in order to 
enhance fainter structures. Apart from some low frequency noise 6 signals
can unambiguously be identified. They are listed in 
Table~2.

\begin{table}
\label{Table: TT Ari periods}
	\centering
	\caption{Significant signals detected in the power spectrum of
                 TT~Ari and their interpretation.}

\begin{tabular}{lll}
\hline
Frequency (d$^{-1}$) & Period (d) & Interpretation \\
\hline
%
$F_1 = \phantom{1}0.9010 (5) $ & $P_1 = 1.1099 (6)$   & $F4 - F3$ \\
$F_2 = \phantom{1}1.8103 (16)$ & $P_2 = 0.5524 (5)$   & $2F_1$    \\
$F_3 = \phantom{1}6.6210 (10)$ & $P_3 = 0.15103 (2)$  & pos.\, superhump \\
$F_4 = \phantom{1}7.5233 (2) $ & $P_4 = 0.132921 (2)$ & neg.\, superhump \\
$F_5 =           13.2393 (20)$ & $P_5 = 0.07553 (1)$  & $2F_3$ \\
$F_6 =           15.0468 (20)$ & $P_6 = 0.066459 (9)$ & $2F_4$ \\
\hline
\end{tabular}
\end{table}

It is easy to interpret the different signals. $F_4$ is by far the 
strongest. The corresponding period being very close to the average
period of the negative superhump there is no doubt that it is due to this 
permanent modulation in TT~Ari. A much fainter signal at $F_3$ has a
period just slightly above the range of periods measured when the positive
superhump appeared between 1997 and 2004 \citep[see table 3 of][]{Bruch19}.
Knowing that the superhump periods are slightly variable, it is close at
hand to identify $F_3$ with TT~Ari's positive superhump. 
$F_1 \approx F_4 - F_3$ is then due to the beat between the negative and positive
superhumps. $F_2$, $F_5$ and $F_6$ are obviously the first overtones of
$F_1$, $F_3$ and $F_4$. The slightly asymmetric waveforms of the superhump
and beat modulations are shown in the bottom frames of Fig.~\ref{ttari}. 

The remarkable feature of the TESS light curve of TT~Ari is thus the
simultaneous appearance of both, negative and positive, superhumps. This
has never been seen before in this system. But in view of the strong
difference in power of the respective power spectrum signals it is doubtful
if the fainter positive superhump would be seen in terrestrial data sets,
even if it were present simultaneously. Thus, both superhumps may have
coexisted previously.

Concluding, I note that the power spectrum of TT~Ari contains a broad
range of enhanced power between 50 and 100~d$^{-1}$ (corresponding roughly
to periods between 15 and 30~min). Signals in this range have often been
seen in the past \citep[see Sect.~3.2 of][and references therein]{Bruch19}
and are generally referred to as quasi-periodic oscillations (QPOs).

\subsection{AC Cnc: A 4.6~d period}
\label{AC Cnc}


AC~Cnc was detected as a variable star by \citet{Kurochkin60}. 
\citet{Kurochkin80} first saw eclipses in the system and determined the
orbital period which was last refined to be 0.300\,477\,47~(4)~d by 
\citet{Thoroughgood04} who also determined the orbital inclination
$i = 75\hochpunkt{o}6 \pm 0\hochpunkt{o}7$.
The system thus belongs to the long period CVs. Slight period changes 
have been analysed by \citet{Qian07}. Soon after the detection of eclipses
the nature of AC~Cnc as a CV was disclosed in spectroscopic observations by 
\citet{Okazaki82}, and in the same year \citet{Downes82} detected the
secondary star in the spectrum, making AC~Cnc one of the few double-lined 
eclipsing variables among CVs and thus enabling the measurement of masses
with a minimum of assumptions \citep[see][]{Schlegel84, Thoroughgood04}.
A limited amount of time resolved photometry of AC~Cnc has been published
by \citet{Yamasaki83}

TESS observed AC~Cnc in two sectors in subsequent time intervals (LC\#3). 
6.5 and 3.5~yr
before this epoch, the system was observed by the Kepler satellite in high
cadence mode as part of the K2 mission (LC\#1 and LC\#2). All light curves
are shown on the same time scale in Fig.~\ref{accnc-lc}. AC~Cnc exhibits
primary and secondary eclipses (see below). The primary eclipses are seen
in the figure as a dense series of excursions to (almost) the same low flux 
level. The secondary eclipses cannot be resolved on the scale of the figure.
The out-of-eclipse flux level shows significant variations on the time 
scale of days. These are not artefacts of the data reduction (see 
Sect.~\ref{Data and data handling}) since they also appear in the PDCSAP 
data.

\begin{figure}
	\includegraphics[width=\columnwidth]{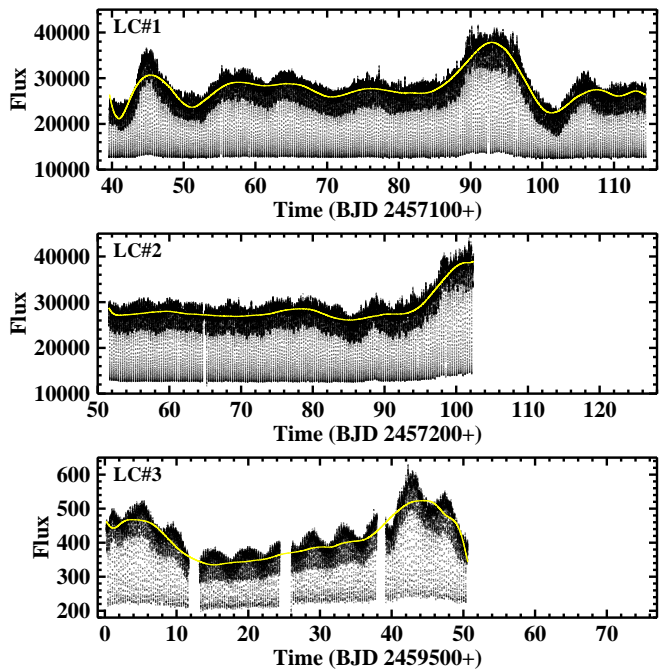}
      \caption[]{Light curves of AC~Cnc observed as part of the Kepler
                K2 mission (LC\#1 and LC\#2) and by TESS (LC\#3), all
                drawn on the same time scale. The yellow graphs represent
                smoothed versions of the light curves after applying a
                Savitzky-Golay filter with a cut-off time scale of 17~d
                (primary and secondary eclipses masked).} 
\label{accnc-lc}
\end{figure}

\begin{figure}
	\includegraphics[width=\columnwidth]{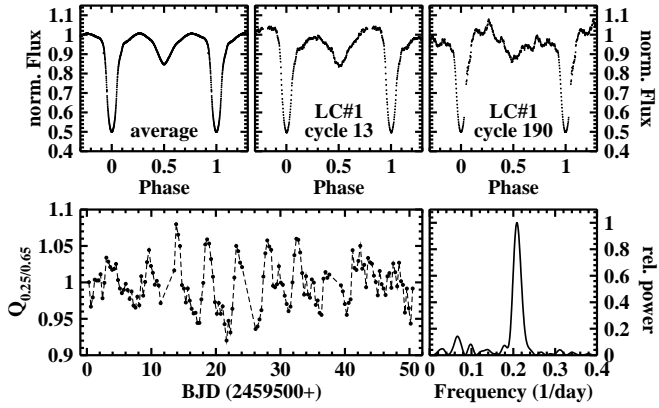}
      \caption[]{{\it Top:} Average waveform (left) of the orbital variations
                of AC~Cnc normalized such that the eclipse minimum has 50\% of 
                the average flux of the two maxima. The middle and right
                frames show examples of individual cycles where the 
                brightness levels of the maxima before and after the
                secondary eclipse are significantly different.
                {\it Bottom:} Quotient of the average normalized flux in
                the phase intervals 0.2-0.3 and 0.7-0.8 for each cycle in LC\#3 
                as a function of time (left) and the corresponding power
                spectrum (right).} 
\label{accnc-wf}
\end{figure}

The average waveform of the orbital variations of AC~Cnc, based on 467
individual cycles, is shown in the upper
left frame of Fig.~\ref{accnc-wf}. It is obvious from Fig.~\ref{accnc-lc}
that the depth of the primary eclipse increases with increasing
system brightness. Therefore the light curves of individual orbital
cycles were normalized before averaging such that the flux at eclipse
minimum is 50\% of the mean flux in the phase intervals 0.2 -- 0.3 and
0.7 -- 0.8. Thus, the average only reflects the shape of the waveform,
not the flux differences. Apart from the primary eclipse it is characterized
by a secondary eclipse at phase 0.5 which interrupts sinusoidal variations
with twice the orbital period and maxima close to phase 0.25 and 0.75.
At the long period of AC~Cnc it is expected (and confirmed by its imprint on
the spectrum of the system) that the secondary star contributes 
non-negligibly to the total light. Considering also the red passbands of the
TESS and Kepler satellites, the
quasi-sinusoidal out-of-eclipse variations are readily interpreted as being 
due to ellipsoidal variations of the late type star. These variations as
well as the secondary eclipse are not seen in earlier terrestrial light
curves \citep[e.g.,][]{Yamasaki83} which were observed at shorter 
wavelengths. 

However, this is not
the whole story. The waveforms of individual cycles, as evidenced by the
examples in the middle and right frames in the top row of 
Fig.~\ref{accnc-wf}, can exhibit a considerable difference in the flux of 
the orbital maxima. These do not occur at random. In the lower left
frame of the figure the quotient $Q_{2.5/7.5}$ of the average normalized flux 
in the phase intervals 0.2-0.3 and 0.7-0.8 of each cycle in LC\#3 is 
plotted as a function of time. $Q_{2.5/7.5}$ oscillates with a period of     
4.79~(3)~d
as indicated by the power spectrum in the lower right frame of the figure.
A similar, albeit fainter, oscillation of $Q_{2.5/7.5}$ with a period of
4.77~(9)~d
is also seen in LC\#2, but not in LC\#1. It thus appears that a wave with a
period just below 5~d modulates the light curves of AC~Cnc.

Finally, I 
note a break in the gradient of the egress of the primary eclipse. Starting
at phase $\approx$ 0.06 it becomes smaller. This feature, seen also in the
eclipse shape of many other eclipsing CVs, indicates a non-axisymmetric 
brightness distribution of the eclipsed body, probably a hot spot at the
disk edge.  

The constancy of the flux level at eclipse minimum in spite of variations
of the total system brightness indicates that these modulations occur in
a part of the system which is totally eclipsed. Only when the brightness
rises much above the average a small part of the extra light evades eclipse,
causing a slight increase of the mid-eclipse flux (e.g., at
BJD~2457190-95 in LC\#1). On the other hand, the minimum flux of the
secondary eclipse is proportional to the average out-of-eclipse flux
(constant of proportionality: 0.9).

\begin{figure}
	\includegraphics[width=\columnwidth]{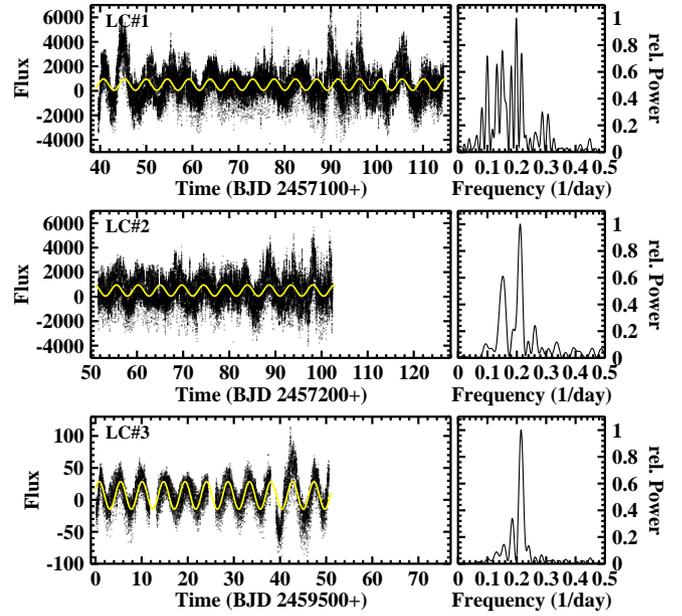}
      \caption[]{{\it Left:} Difference between the original light curves
                of AC~Cnc (primary and secondary eclipses masked) and their 
                smoothed versions (yellow graphs in Fig.~\ref{accnc-lc}).
                {\it Right:} Low frequency power spectra of the light
                curves. The yellow graphs on the left are least squares
                sine fits with the period fixed to the period of the 
                dominant power
                spectrum signal (second highest peak in the case of LC\#1).} 
\label{accnc-lf}
\end{figure}

In particular in LC\#3 the variations on time scales of days appear to
show some regularity. To investigate this issue further, a Savitzky-Golay
filter with a cut-off time scale of 17~days was applied to all light
curves (masking both, primary and secondary eclipses), resulting in the
smoothed light curves shown in yellow in Fig.~\ref{accnc-lc}. The difference
between the original (again masking both eclipses) and the smoothed curves
is shown in Fig.~\ref{accnc-lf} (left) together with their low frequency
power spectra (right). The power spectra of LC\#2 and LC\#3 clearly
indicate the presence of a variation with a period of 
4.73~(5) and
4.66~(2)~d,
respectively; identical within the error margin. The power spectrum of LC\#1
contains several peaks, one of the strongest very close in 
frequency to the peaks seen in the other light curves and corresponding to 
a period of 
4.63~(2)~d.
Least squares
sine fits with periods fixed to the respective values are shown in yellow
in Fig.~\ref{accnc-lf}. Is this the manifestation of the beat of orbital
variations with some other periodicity? This is also suggested by
the variations of $Q_{2.5/7.5}$ in LC\#2 and LC\#3 which occur at the same
periods. At higher frequencies, only the power spectrum of
LC\#3 contains a faint signal at 3.4305~(15)~d$^{-1}$,
indicating the presence of a negative superhump with a period of
0.2924~(1)~d.
Otherwise, the power spectra of all light curves are strongly dominated
by the orbital frequency and its overtones, which can clearly be 
identified up to the Nyquist frequency in the Kepler data (222$^{th}$ 
overtone!). No other significant signals are detected. 

The orbital period of AC~Cnc was last discussed in detail by \citet{Qian07}
who found a secular period decrease and suspected a superposed cyclic 
variation with a period of 16.2~yr. The current data permit to extend the
time base for period determination significantly. Each of the light curves
contain between 151 and 249 eclipses. Instead of measuring individual eclipse 
timings the light curves were folded on the orbital period choosing as
epoch for the zero point of phase an instant approximately in the middle of
the light curve such that the centre of the primary eclipse coincides with
phase 0. This yields the 3 high precision eclipse epochs for the three
light curves listed in 
Table~3.

\begin{table}
\label{Table: eclipse epochs}
	\centering
	\caption{Representative eclipse epochs.}

\begin{tabular}{llll}
\hline
Star   & Light curve & Epoch (BJD) & Cycle number$^1$ \\
\hline
AC Cnc & LC\#1 & 2457179.2850 & 42895 \\
       & LC\#2 & 2458281.1357 & 46562 \\
       & LC\#3 & 2459520.0024 & 50685 \\ [1ex]
TV Col & LC\#1 & 2458461.1377 & 52749 \\
       & LC\#2 & 2459199.1453 & 55973 \\ [1ex]
AQ Men & LC\#1 & 2458340.2116 & 0     \\
       & LC\#2 & 2458352.1133 & 791   \\
       & LC\#3 & 2458529.0722 & 1335  \\
       & LC\#4 (part 1) & 2458611.1237 & 1915 \\
       & LC\#4 (part 2) & 2458638.0025 & 2105 \\
       & LC\#4 (part 3) & 2458669.1256 & 2325 \\
       & LC\#5 (part 1) & 2459050.0996 & 5018 \\
       & LC\#5 (part 2) & 2459073.0176 & 5180 \\
       & LC\#6 & 2459156.0597 & 5767 \\
       & LC\#7 & 2459266.1223 & 6545 \\
       & LC\#8 (part 1) & 2459348.0324 & 7124 \\
       & LC\#8 (part 2) & 2459373.0721 & 7301 \\ [1ex]
\hline
\end{tabular}
$^1$ cycle count convention according to: 
\citet{Qian07} (AC~Cnc);
\citet{Augusteijn94} (TV~Col);
this work (AQ~Men)
\end{table}

Combining these new eclipse epochs with those listed in Table~1 of 
\citet{Qian07}, I updated the orbital ephemeris of AC~Cnc, adopting
a linear as well as quadratic model. Following \citet{Qian07} weights
of 1 and 5 were assigned to epochs based on photographic and 
photoelectric/CCD observations, respectively. 
The three new data points, all representing
a large number of eclipses, arbitrarily got weight 10. Quadratic
ephemeris clearly describe the eclipse epochs better than linear ones:
\begin{eqnarray}
\label{AC Cnc ephemeris}
T_{\rm min} & = & BJD 2444290.3076\, (5) \\ \nonumber
          &   & + 0.300\,477\,52\, (2) \times E \\ \nonumber
          &   & - 3.2 \,(4)\, 10^{-12} \times E^2
\end{eqnarray}
Here, the errors are the formal fit errors. This result confirms the
secular decrease of the orbital period of AC~Cnc, although at a rate
smaller (but still within the error limits) than reported
by \citet{Qian07}.

\begin{figure}
	\includegraphics[width=\columnwidth]{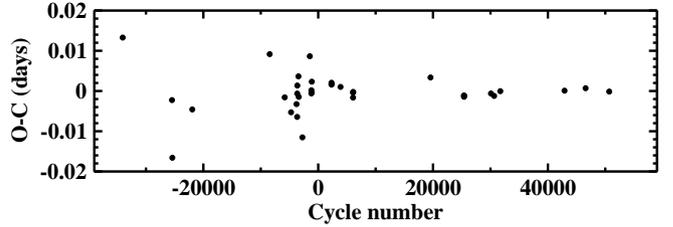}
      \caption[]{$O-C$ diagram of AC~Cnc based on the ephemeris 
                 given in Eq.~\ref{AC Cnc ephemeris}.}
\label{accnc-oc}
\end{figure}

The $O-C$ diagram is shown in Fig.~\ref{accnc-oc}. As expected the scatter 
is much higher for the photographically determined eclipse epochs (all data 
before cycle 0). At later epochs $O-C$ is almost constant and close to 0.
In particular, the presence of cyclic variations with a period of 16.2~y, 
suspected by \citet{Qian07} based on the photoelectric/CCD $O-C$ values and 
depending decisively on the single data point close to cycle 20\,000, is 
not only not sustained by the present results but definitely ruled out.
Thus, the conclusions drawn by them from the cyclic variations are obsolete.

\subsection{QU Car: Finally a more accurate orbital period}

QU~Car was discovered as an irregular variable by \citet{Stephenson68}
and classified as a cataclysmic variable by \citet{Schild69}. At an average
magnitude slightly fainter than 11~mag it is one of the brightest
CVs in the sky. Currently considered to be a novalike variable,
\citet{Kafka08} also proposed a classification as V~Sge star, i.e., 
a system with characteristics similar to supersoft X-ray sources
\citep{Steiner98}. Despite its high brightness QU~Car is a much less
well studied system than many other much fainter CVs. Even the most
basic of all binary parameters, the orbital period, is only know with a
rather low precision. It
was first determined spectroscopically by \citet{Gilliland82}
as $10.9 \pm 0.3$~h. While \citet{Kafka08} did not see this period in
their observations, it was later confirmed at an equally low precision
by \citet{Oliveira14}. These authors
also find a photometric modulation with the same period, but
only when the flickering activity is reduced and the system hovers
at a slightly lower magnitude level than normal.

\begin{figure}
	\includegraphics[width=\columnwidth]{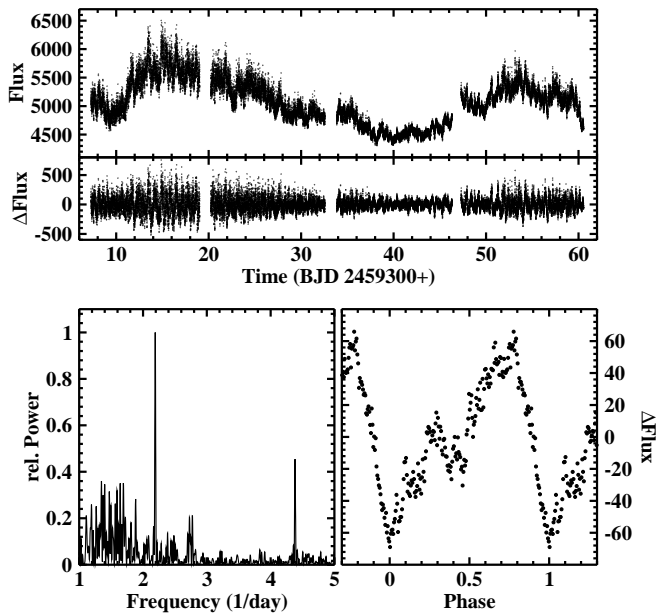}
      \caption[]{{\it Top:} Light curve of QU~Car (upper frame) and
                the same after subtraction of variations
                on time scales longer than 2 days (lower frame).
                {\it Bottom:} Low frequency part of the power spectrum
                of QU Car, dominated by signals at the orbital frequency and
                the first overtone (left) and average waveform of orbital 
                variations of QU~Car (right).}
\label{qucar-lc}
\end{figure}

The TESS light curve of QU~Car is shown in the upper frame of 
Fig.~\ref{qucar-lc}. It is characterized by gradual variations with
two well expressed maxima, separated by $\approx 38$~d (the time base
is evidently too short to affirm if they are periodic).
This modulation is superposed
upon variations on shorter time scales which are markedly stronger close
to the maxima of the long term variations. While the eye does not clearly
detect periodic variations on short time scale, this becomes different
when modulations on time scales 2~d or longer are subtracted (second frame
of Fig.~\ref{qucar-lc}). Regular variations are now unambiguously seen.
The power spectrum of this light curve (lower left frame of 
Fig.~\ref{qucar-lc}) reveals irregular variations at frequencies $<2$~d$^{-1}$
(very low frequencies are suppressed by the filtering process), and is
dominated by a strong signal at 
$F=2.192~(1)$~d$^{-1}$ 
($P=0.4563\, (2)\, {\rm d} = 10.951\, (5)\, {\rm h}$) 
and its first overtone. This is obviously the orbital frequency and thus 
confirms and significantly improves the
period found by \citet{Gilliland82} and \citet{Oliveira14}.

Folding the data on the orbital period in order to determine the waveform
and a fiducial structure to define a zero point of phase is complicated by
(i) strong flickering and (ii) the strongly varying amplitude of the orbital 
variations (second frame of Fig.~\ref{qucar-lc}). Assuming that the latter
does not alter significantly the shape of the waveform this problem can be
resolved by a proper normalization. For this purpose, I apply a spline fit 
to a manually defined approximate upper envelope of the orbital variations
in the second frame of Fig.~\ref{qucar-lc} and divide the light curve by the 
spline. Folding
the results on the orbital period and binning them in phase intervals of
width 0.005 then results in the average waveform displayed in the lower right
frame of Fig.~\ref{qucar-lc}. It exhibits a gradual rise from a minimum 
(defined as phase 0) to a maximum at phase 0.75, and then a much quicker
drop to the next minimum. This behaviour is quite similar to that observed
by \citet{Oliveira14} (see their fig.~11), meaning that QU~Car did not change
its waveform in the $\approx$9 years between the data sets.


With the orbital minimum defined as zero point of phase, it is now possible
to define ephemeris:
\begin{equation}
T_{\rm min} = BJD\, 2459342.0721 (5) + 0.4563 (2) \times E
\end{equation}
Here, the error of the epoch 
corresponds to a shift of 0.01 in phase of the orbital 
minimum. 

Apart from the orbital signal, fainter isolated peaks in the power spectrum
may hint at other persistent periodic modulations in the light curve. The 
most obvious of these is a pair of signals at 
2.728~(2)~d$^{-1}$ and 2.774~(3)~d$^{-1}$
(corresponding to periods of 
8.796~(8)~h and 8.652~(7)~h; see Fig.~\ref{qucar-lc}). 
These frequencies have no
obvious relation to each other or to the orbital frequency. Time resolved
power spectra revealed that the signals are mainly caused by variations 
during a few days close to the end of the observations, while at other 
intervals they are at most marginally present. Therefore, the evidence
for them to be persistent is weak. Similarly, time resolved power spectra
show that a candidate signal at higher frequencies (38.25~d$^{-1}$) is not
significant.

\subsection{V504 Cen: A supraorbital period}

Originally classified as a R~CrB star \citep{Kholopov85}, based on its 
spectrum \citet{Kilkenny89} identify V504~Cen as a cataclysmic variable.
The photometric behaviour made them suspect that V504~Cen belongs to the
VZ~Scl class. This was confirmed by \citet{Kato03} and \citet{Greiner10}
and even more impressively by the long term AAVSO light curve which --
after a long sojourn in a high state -- exhibits frequent and rapid 
successions of high and low states in recent years. \citet{Greiner10}
detected photometric variations in optical light, as well as in X-rays, 
and radial velocity variations which permitted them to determine an orbital 
period of 0.175\,566\,55~d.

\begin{figure}
	\includegraphics[width=\columnwidth]{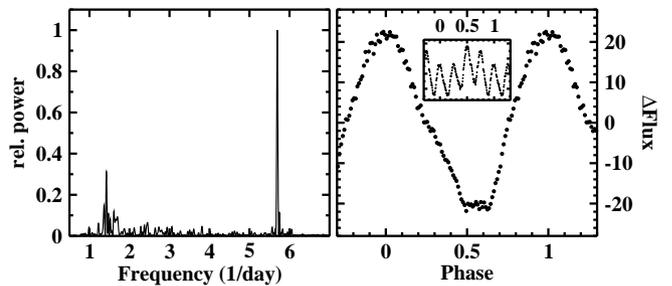}
      \caption[]{{\it Left:} Power spectrum of the light curve of
                V504~Cen.
                {\it Right:} Light curve of V504~Cen folded on the
                orbital period and on 4 times the orbital period (insert).}
\label{v504cen-ps}
\end{figure}

TESS observed V504~Cen in a short intermediate state between two low
states, about 1~mag below the normal high state. The light curves exhibits
some irregular variations on the time scale of days which were subtracted
to yield the power spectrum reproduced in the left frame of 
Fig.~\ref{v504cen-ps}. It is dominated by a strong signal at the orbital 
frequency. At higher frequencies, only its first overtone is faintly 
present. The phase folded light curve in the right frame of the figure
is characterized by a strong hump. Interestingly, it contains a constant
interval between phase 0.485 and 0.635 (where the hump maximum defines
phase 0). This gives the impression that the light source responsible for
the orbital variations is temporarily blocked from view. In spite of having
been observed at a similar wavelength this waveform is
different from that obtained by \citet{Greiner10} (see their figure 2)
when V504~Cen was in a genuine high state, and which has a flat topped maximum
and a sharp minimum.

As an unusual feature the power spectrum of V504~Cen indicates the presence
of another periodic signal at exactly 1/4 of the orbital frequency. The
corresponding additional modulation of the brightness of the system at 4
times the orbital period is evident in the light curve, folded on this
period, shown in the insert in the right frame of Fig.~\ref{v504cen-ps}.
Every third and fourth maximum is significantly brighter than the first
and second.
This odd behaviour cannot be explained by the presence of variations with
independent periods (such as superhumps). I am not aware of other
systems showing anything similar.

\subsection{TV Col: The spin period finally detected in optical light}

The intermediate polar TV~Col was identified as the optical counterpart
of the hard X-ray source 2H~0526-328 by \citet{Charles79}. It
exhibits complicated photometric variations which
include 3 periods observed in optical light, in addition to another
period only seen in X-rays. The detection by \citet{Schrijver85}
of a 1938~s X-ray period, also claimed to have been seen in the $UBV$
bands by \citet{Bonnet-Bidaud85}, was later found to be ambiguous,
the true period being 1911~s \citep{Schrijver87}. A more accurate
value of $1909.67 \pm 2.5$~s is provided in table 2 of \citet{Rana04}.
This period is interpreted
as the rotation period of the white dwarf, giving rise to the classification
of TV~Col as intermediate polar. It was
never observed in the optical. Instead, optical variations occur at
the spectroscopic period, considered to be orbital, first measured by 
\citet{Hutchings81} and later refined to be 0.228\,600\,3~(2)~d by 
\citet{Hellier93}. Another shorter period modulation at 0.213\,04~d, thought to
be due to a negative superhump, is also often
seen together with its beat with the orbit at 4.024~d 
\citep{Motch81, Hutchings81, Barrett88} and may be slightly unstable
\citep{Augusteijn94}. \citet{Hellier91} found shallow partial eclipses
in the light curve of TV~Col. Finally, \citet{Retter03} claim to have
seen a positive superhump with a period of 0.264~d.

\begin{figure}
	\includegraphics[width=\columnwidth]{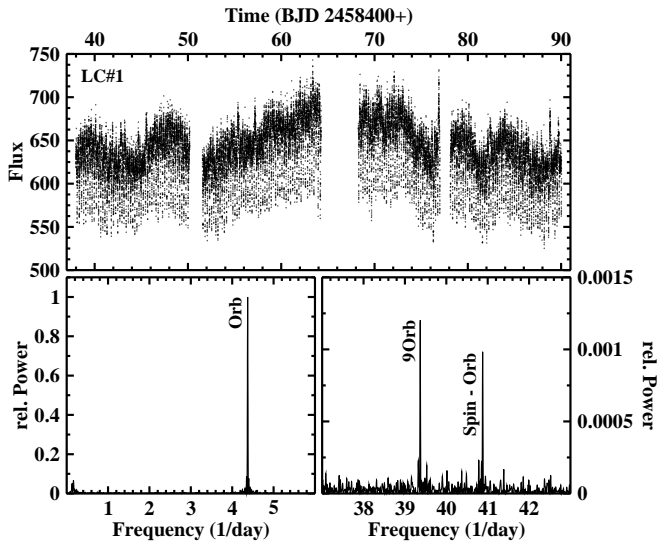}
      \caption[]{{\it Top:} Light curve LC\#1 of TV~Col.  {\it Bottom left:}
                Low-frequency part of the power spectrum of LC\#1 containing
                a strong signal at the frequency of the orbital variations.
                {\it Bottom right:} The same power spectrum at higher 
                frequencies showing apart from an overtone of the orbital
                frequency a signal due to the beat between the orbital
                variations and variations caused by the spin of the white
                dwarf.}
\label{tvcol-lc1}
\end{figure}

\begin{figure}
	\includegraphics[width=\columnwidth]{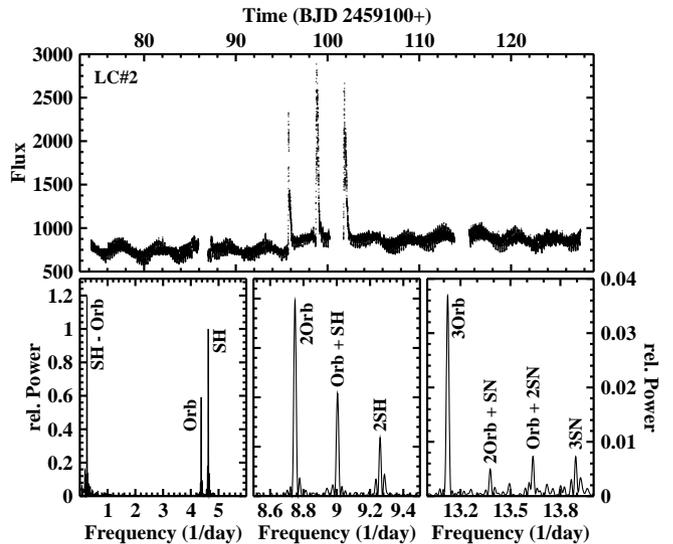}
      \caption[]{{\it Top:} Light curve of LC\#2 of TV~Col containing
                three short-term flares.
                {\it Bottom left:} Low frequency part of the power spectrum
                of LC\#2, containing signals due to the orbital and 
                negative superhump variations, as well as to the beat 
                between them. {\it Bottom middle and right:} The same power
                spectrum at higher frequencies, showing examples of signals
                at linear combinations of the orbital and the superhump
                frequencies.}
\label{tvcol-lc2}
\end{figure}

Two TESS light curves of TV~Col, separated by about 2 yr are avaialble
and shown in the upper frames of 
Figs.~\ref{tvcol-lc1} and \ref{tvcol-lc2}. LC\#2 contains three short term
outbursts (at BJD 2459194, 2459198 and 2459201) such as those observed 
by \citet{Szkody84}, \citet{Augusteijn94} and \citet{Hudec05}. They are 
ignored in the following analysis. It is obvious that the temporal
behaviour of TV~Col is quite different in the two light curves.

The low frequency part of the power spectrum of LC\#1 is shown in the
lower left frame of Fig.~\ref{tvcol-lc1}. Apart from some very low
frequency noise it contains only a single strong signal at the orbital
frequency. No trace of the often seen superhump period at
$\approx$4.63~d$^{-1}$ or its beat with the orbital frequency is present.
Thus, the superhump was definitely absent at the epoch of this light
curve. A never before seen feature is revealed in the lower right frame
of the figure which contains a small section of the power spectrum at
higher frequencies. Apart from a high overtone of the orbital frequency
a signal at 
40.878~(2)~d$^{-1}$
is present which within the error margin is equal to 
$F_{\rm spin} - F_{\rm orb} = 40.87 (6)$~d$^{-1}$. 
It is thus the beat
between the white dwarf spin and the orbital period (i.e., the orbital
sideband of the spin), a signal which is 
commonly seen in the light curves of intermediate polars instead of the
spin period itself. This is the first time that a manifestation of the 
white dwarf rotation period is seen in optical light of TV~Col. The
signal is also present in the power spectrum of LC\#2.

Otherwise, LC\#2 behaves quite differently. The light curve contains
obvious periodic variations on the time scale of days which immediately
suggest to be due to the beat between the orbital and the commonly 
observed superhump period of TV~Col. This is confirmed by the power
spectrum, the low frequency part of which is reproduced in the lower
left frame of Fig.~\ref{tvcol-lc2}. The superhump is again active at
4.6307~(3)~d$^{-1}$ ($P = 0.215\,95\, (1)$~d)
and generates a stronger signal than the orbital variations. But the
beat between the two at
0.2548~(4)~d$^{-1}$ ($P = 3.894\, (5)$)
contains even more power. Linear combinations of
$F_{\rm SH}$ and $F_{\rm orb}$ give rise to additional signals, as 
exemplified in the two other frames in the lower part of the figure.

\begin{figure}
	\includegraphics[width=\columnwidth]{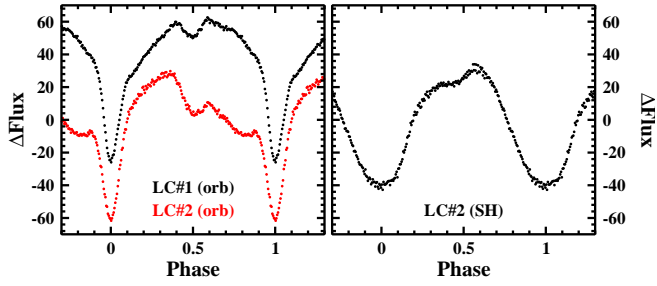}
      \caption[]{{\it Left:} Waveform of the orbital variations
                of TV~Col of LC\#1 (black, shifted vertically for clarity) and
                LC\#2 (red).
                {\it Right:} Waveform of the negative superhump
                observed in LC\#2.}
\label{tvcol-wf}
\end{figure}

The occurrence of eclipses in TV~Col, not easily discerned in individual
cycles in terrestrial observations, is impressively confirmed when
the light curves are folded on the orbital period. The waveform of the
orbital variations is shown in the left frame of Fig.~\ref{tvcol-wf}. 
For the first time, a secondary eclipse is also documented. While the
waveform derived from LC\#1 is approximately symmetrical, this is not
the case for LC\#2. It appears that, not surprisingly, the presence
of the superhump light source implies in a significant change in the
structure of the accretion flow in TV~Col. To complete this issue
the waveform of the superhump is shown in the right frame of 
Fig.~\ref{tvcol-wf}.  

Using the same technique as in the case of AC~Cnc (see Sect.~\ref{AC Cnc})
to define representative eclipse epochs for the two light curves of TV~Col
and combining them with previous measurements it is possible to improve the 
orbital period. The ephemeris of \citet{Hellier93} or \citet{Augusteijn94}
are sufficiently accurate for an 
unambiguous cycle count in spite of the long time interval between their 
observations and the TESS data. Eclipse epochs and cycle numbers 
derived from the present light curves are listed in 
Table~3. Combining eclipse epochs listed
by \citet{Hellier93} and \citet{Augusteijn94}
(assigning weight 1) with the new ones (weight 10) then yields the
following ephemeris for the eclipses of TV~Col:
\begin{equation}
T_{\rm ecl} = BJD\, 2446403.7114\, (5) + 0.228\,600\,10\, (2) \times E 
\end{equation} 
where the errors are the formal errors of a linear fit to the cycle number
vs.\ eclipse epoch relation. \citet{Hellier93} also considered quadratic
ephemeris for the eclipses in TV~Col. While the quadratic term is formally 
highly significant it depends on a single data point. Therefore,
\citet{Hellier93} regards this solution suspect. The present additional
eclipse timings do no warrant quadratic ephemeris.

\subsection{DM Gem: Orbital and superhumps periods: which is which?}

DM~Gem is an old nova which erupted in 1903. Very few details of the
underlying binary system are known. \citet{Lipkin00} observed quasi-periodic
oscillations which they interpreted as a combination of two distinct signals
with closely spaced periods near $\sim$2.95~h. Variations at the same 
period were also seen in photometric observations of \citet{Rodriguez-Gil05}.
They consider them to be probably linked to the orbital period. 
The latter authors also claim the presence of 
short-term variations on the time scale of $\sim$20~min.

The TESS SAP light curve of DM~Gem  contain
some features such as systematically different flux levels in adjacent
parts of the light curves, significant linear gradients or even apparent
``outbursts'' which are clearly artefacts. All of these are absent in the
PDCSAP data. Therefore, in this case, I prefer to use the latter.

\begin{figure}
	\includegraphics[width=\columnwidth]{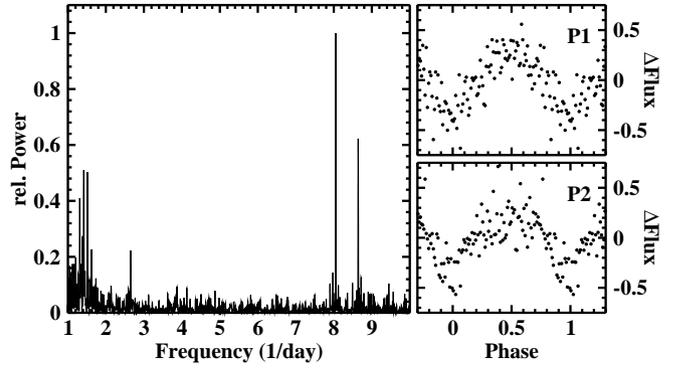}
      \caption[]{{\it Left:} Power spectrum of the light curve of
                DM~Gem
                {\it Right:} Light curve of DM~Gem folded on the
                periods $P_1$ (top) and $P_2$ (bottom), binned in phase
                intervals of width 0.01.}
\label{dmgem-ps}
\end{figure}

The power spectrum of the light curve is shown in the left frame of
Fig.~\ref{dmgem-ps}. There is the usual forest of peaks at low frequencies
caused by residual irregular variations on longer time scales. But dominating
are two peaks at
$F_1 = 8.0495\, (8)\, {\rm d}^{-1}$ ($P_1 = 0.124\,23\, (1)\, {\rm d}$) and
$F_2 = 8.0430\, (10)\, {\rm d}^{-1}$ ($P_2 = 0.115\,70\, (1)\, {\rm d}$), 
the latter being somewhat
fainter. The amplitude of these variations is quite small, such that the
folded light curves, even after binning, are rather noisy, as seen in the
two right hand frames of the figure. In both cases the waveform is sinusoidal.
A time resolved power spectrum (not 
shown) reveals that the strength of the signal is quite variable and may
remain below the detection level for intervals of days. This is particularly
true for the $P_2$ signal.

The minima of the $P_1$ and $P_2$ variations can be described by the following
ephemeris:
\begin{eqnarray}
P_1: & T_{\rm min} = BJD\, 2459514.11\, (1) + 0.124\,23\, (1) \times E \\
P_2: & T_{\rm min} = BJD\, 2459514.04\,(1) + 0.115\,70\, (1) \times E 
\end{eqnarray}
In view of the scatter in the waveforms the errors of the epochs are defined 
allowing for an uncertainty of the minimum epoch of 0.1 times the period.

Within the error limits quoted by \citet{Rodriguez-Gil05} the period seen
in their light curves is identical to $P_1$. If this is the orbital period
of DM~Gem, it is close at hand to suspect $P_2$ to be the period of a
negative superhump. Or should it be vice versa, $P_2$ being the orbital
and $P_1$ a positive superhump period? The period difference is 6.9\%. 
The average differences for negative and positive superhumps calculated
from table~4 of \citet{Yang17} is $3.9 \pm 1.8\%$ and $5.1 \pm 2.4\%$,
respectively. This may indicate a slight preference for $P_1$ to be due
to a positive superhump,
but this is far from conclusive. To settle this question, a spectroscopic
determination of the orbital period is required. 

Apart from the $P_1$ and $P_2$ signals and the forrest at low frequencies a
lonesame peak at 
2.650~(2)~d$^{-1}$ ($P = 0.3774\, (2)$~d) 
appears to have some significance. This period being very close to $3P_1$
may or may not be
a coincidence. I cannot offer a viable explanation for this signal. At
higher frequencies there are no indications for coherent or quasi-periodic
oscillations such as those seen by \citet{Rodriguez-Gil05} at $\sim$20~min.

\subsection{RZ Gru: The orbital period and a superhump?}

RZ~Gru is a little studied novalike variable of the UX~UMa subtype.
Originally discovered as a variable star by \citet{Hoffmeister49}
its nature as a cataclysmic variable was first suspected by
\citet{Kelly81} and then confirmed by \citet{Stickland84}.
The orbital period is uncertain. In a short notice \citet{Tappert98}
quote two possible values based on spectroscopic observations on Sep.\ 29
and Oct.\ 2, 1995 (0.360~d and 0.408~d) without indicating error
limits.

\begin{figure}
	\includegraphics[width=\columnwidth]{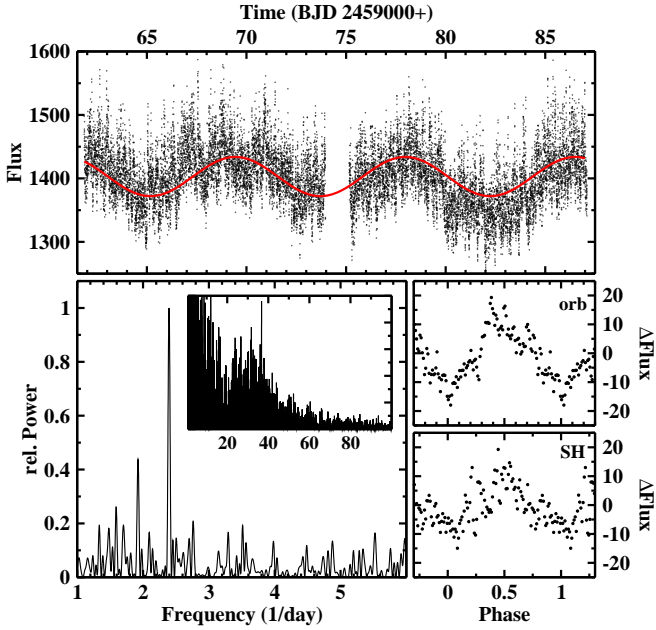}
      \caption[]{{\it Top:} Light curve of RZ~Gru. The red curve
                is a least squares sine fit to the data with period of
                8.53~d. {\it Bottom left:} Power spectrum of the light 
                curve of RZ~Gru. The insert shows a wider frequency range,
                containing a broad range of QPOs.
                {\it Bottom right:} Light curve of RZ~Gru folded on the
                orbital (top) and the superhump (bottom) period, binned in phase
                intervals of width 0.01.}
\label{rzgru-lc}
\end{figure}

The TESS light curve of RZ~Gru is plotted in the upper frame of 
Fig.~\ref{rzgru-lc}. On longer time scales it is characterized by
a nearly sinusoidal variation. This is preserved in the PDCSAP data and
is therefore probably real. A formal least squares fit (red graph in
the figure) yields a period of 8.53~d. The power spectrum 
of the original light curve has a strong peak corresponding to a 
very similar period of 
8.55~(13)~d. Of course, more cycles have to be covered to be certain that
this apparent period is persistent.

The power spectrum after subtracting variations on time scales $>2$~d is
shown in the lower left frame of Fig.~\ref{rzgru-lc}. It is dominated by 
a maximum corresponding to a period of
$P_1 = 0.4175\,(8)$~d.
This is quite close
to one of the two options quoted by \citet{Tappert98} for the orbital period.
They did not quote errors. But their observations were obtained in two nights
spanning a time base of at most $\approx$3.5~d or $\approx$8.4 cycles. A
period error equal to the difference between their period and $P_1$ would
therefore lead at most to a phase shift of 0.20. Depending on the unknown
quantity and quality of the spectroscopic data a period error leading to a
phase shift of this magnitude is not unconceivable. Therefore, I will assume 
$P_1 = P_{\rm orb}$. 

Another peak at a slightly lower frequency of 
$F_2 = 1.924~(7)$~d$^{-1}$  
corresponds to a period of
$P_2 = 0.520\, (2)$~d.  
I tentatively identify it as due to a (positive) superhump. 
In contrast to other systems, the power spectrum of the original data does
not contain a significant peak at the beat frequency between orbital and
superhump frequencies.
The light curve, folded on $P_{\rm orb}$ and $P_2 = P_{\rm SH}$ and binned into
intervals of 0.01 in phase, is plotted in the lower right frames of 
Fig.~\ref{rzgru-lc}.
The minima of the orbital variations can be described by the
ephemeries:
\begin{equation}
T_{\rm min} = BJD\, 2459074.08\, (4) + 0.4175\, (8) \times E \\
\end{equation}
For the same reason as in the case of DM~Gem, the error of the minimum 
epoch allows for an uncertainty of 0.1 times the period.

The beat frequency between the orbital and superhump period at 2.12~(4) is not
seen in the power spectrum. Its period is also much smaller than the period of
the strong 8.55~d modulation in the light curve. But is it a coincidence
that within 0.5 times the propagated formal error the latter is an 
integer multiple (i.e., four times) of the former? A similar case has
been observed by \citet{Bruch18} in V603~Aql, the brightness of which is
modulated at exactly twice the beat period between orbit and superhump.

A final interesting feature is an enhancement of power at frequencies 
between roughly 20 -- 65~d$^{-1}$ (see insert in the lower left frame of
Fig.~\ref{rzgru-lc}) which may be considered a broad range of QPOs. A
time resolved power spectrum reveals no persistent features in this range.
This distribution of power 
is different from that of most other CVs where red noise caused
by flickering leads to monotonous decrease of power towards high frequencies
\citep[see, e.g.,][]{Bruch22}.

\subsection{V533 Her: Superhumps with period variations}

V533~Her is the remnant of a bright nova which exploded in 1963. The quiescent 
system gained some attention when \citet{Patterson79} observed coherent 
oscillations with a period of 63.3~s which appeared to indicate that 
the system belongs to the rapidly rotating intermediate polars of the DQ~Her 
subclass. However, later observations by \citet{Robinson83}
showed that these oscillations had disappeared, casting doubt on the
magnetic rotator model for V533~Her. Unfortunately, the time resolution of
the TESS data is not sufficient to resolve the corresponding period range and
to investigate this issue further. A spectroscopic orbital period of
$P_{\rm orb} = 0.146\,885\,1\, (3)$~d or 
$P_{\rm orb} = 0.147\,374\,3\, (3)$~d was measured by \citet{Thorstensen00}.
A slightly shorter photometric period of 0.142\,89~(2)~d was observed by 
\citet{McQuillin12} in SuperWASP data. It may be related to a negative 
superhump. 

TESS observed V533~Her during two time intervals separated by about one
year. After submission of the original version of this study
\citet{Leichty22} published a small paper on V533~Her based on the same 
TESS data. They used techniques similar to those applied here, but the 
approach for the interpretation of the observational results is slightly 
different.

\begin{figure}
	\includegraphics[width=\columnwidth]{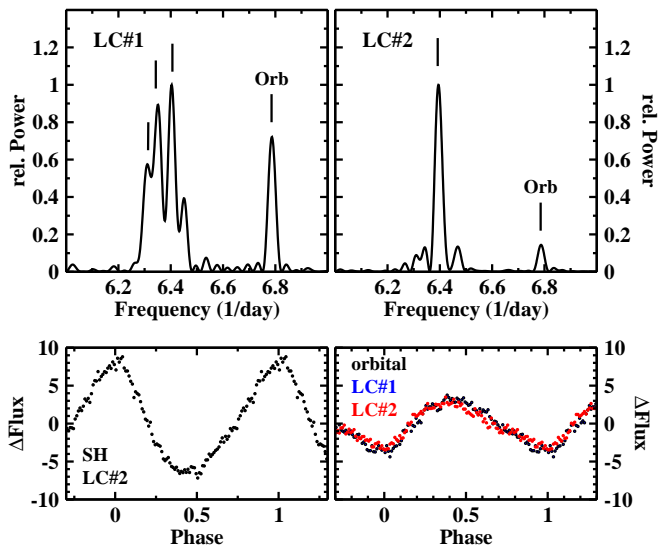}
      \caption[]{{\it Top:} Power spectra of the two light curves of V533~Her
                 in a narrow frequency range around the dominating signals.
                 {\it Bottom:} Waveforms of the superhump variations in LC\#2 
                 of V533~Her (left) and the orbital modulation in LC\#1 and 
                 LC\#2, plotted on the same scale.}
\label{v533her-ps}
\end{figure}

The power spectra of the light curves are dominated by two signals between
6 and 5~d$^{-1}$ as shown in upper frames of Fig.~\ref{v533her-ps}. The
higher frequency peaks can readily be identified as being due to the longer 
of the two candidate orbital periods of \citet{Thorstensen00} (see also
\citet{Leichty22}).


\begin{figure}
	\includegraphics[width=\columnwidth]{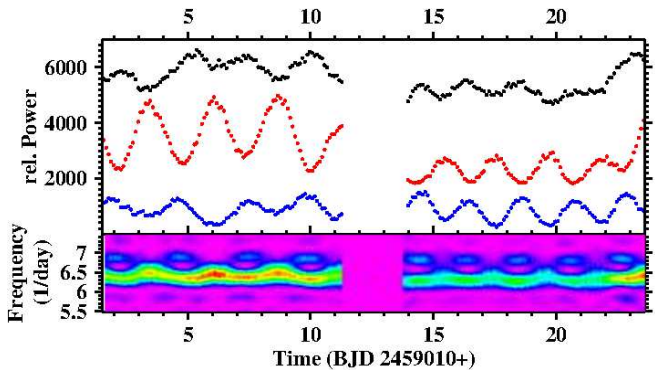}
      \caption[]{Properties of the power spectrum of LC\#1 of V533~Her as
                 a function of time. The lower frame 
                 shows the power on a colour scale as a function of 
                 time and frequency. The time interval affected by the gap 
                 between the two parts of the light curve has been masked.
                 The upper frame contain the power
                 integrated over three small frequency intervals (shifted
                 vertically by an arbitrary amount for clarity). 
                 For details, see text.}
\label{v533her-stacked}
\end{figure}

In agreement with \citet{Leichty22} I interpret the second signal 
as a positive superhump. In LC\#1 it consists of several overlapping peaks 
at frequencies (collectively referred to here as $F_{\rm SH}$) 
between $\approx$ 6.25 and 6.50~d$^{-1}$ in LC\#1, but it has a simpler
structure with a dominating peak at
6.3947~(7)~d$^{-1}$ 
in LC\#2. In order
to understand this behaviour a time resolved power spectrum of LC\#1 was
calculated, using a sliding window of width 2.6~d.
The result is plotted as a function of time and frequency
in the bottom frame of Fig.~\ref{v533her-stacked} (note that the
difference in appearance of the corresponding power spectrum of
\citet{Leichty22} is exclusively due to the larger window used by them). 
The orbital signal
appears and disappears with time in a regular pattern. The power of $F_{\rm SH}$
is modulated similarly in time, but additionally in frequency, being
shifted to higher frequencies whenever the orbital signal disappears.
This behaviour suggests to be due to the interference of two perodic
signals with different amplitudes. This is confirmed by tests with
artificial light curves. A similar effect is also clearly seen in the
power spectrum of \citet{Leichty22} where it manifests itself as wiggles 
in both, the orbital and the superhump frequencies which obviously must
not be interpreted as real variations in the former, and probably also not
in the latter.


But there is more to be extracted from the time resolved power spectrum.
In the upper frame of Fig.~\ref{v533her-stacked} the integrated power in
three small frequency intervals (from top to bottom: 6.20 -- 6.44~d$^{-1}$;
 6.46 -- 6.70~d$^{-1}$ and 6.72 -- 7.04~d$^{-1}$) are plotted as a function
of time (the graphs corresponding to the first two intervals have been shifted
upwards by an arbitrary amount for clarity). The regular pattern seen in the
time resolved power spectrum repeats, but the period is longer before the
light curve gap and shorter afterwards. It appears that the change
occurred within a short time interval. Interpreting the periods as the 
beat between $P_{\rm orb}$ and $P_{\rm SH}$ explains the main peaks of the 
$F_{\rm SH}$ complex marked in Fig.~\ref{v533her-ps}, as is detailed in 
Table~4. This leaves only the third highest 
frequency peak in the $F_{\rm SH}$ complex without an explanation.

\begin{table}
\label{v533her-tab}
	\centering
	\caption{Beat frequencies and periods between $P_{\rm orb}$ and
                 $P_2$, and calculated values for 
                 $F_2 = F_{\rm orb} - F_{\rm beat}$ and $P_2$ based
                 on LC\#1 of V533~Her and its parts LC\#11 (before the gap)
                 and LC\#12 (after the gap). The time unit is the day.}

\begin{tabular}{llll}
\hline

     & LC\#11 & LC\#12 & LC\#11 + LC\#12 \\
\hline

$F_{\rm beat}$ & 0.378 (1)   & 0.472 (3)   & 0.443 (2)\\
$P_{\rm beat}$ & 2.644 (9)   & 2.12 (1)    & 2.26 (1) \\
$F_2$        & 6.407 (1)   & 6.314 (2)   & 6.343 (2) \\
$P_2$        & 0.15607 (3) & 0.15838 (7) & 0.15766 (5) \\
\hline
\end{tabular}
\end{table}

Considering the simpler structure of $F_{\rm SH}$ in LC\#2 a similar excersize 
for this light curve does not yield surprises. The time resolved power spectrum
contains the same beat pattern, but this time the beat frequency does not
change significantly over time. It yields a frequency of
6.392~(2)~d$^{-1}$ ($P_2 = 0.156\,45\,(5)$~d, marked in the right 
frame of Fig.~\ref{v533her-ps}), consistent with $F_{\rm SH}$.

The orbital and superhump waveforms (the latter only for LC\#2 because
of its period variation in LC\#1), binned in phase intervals of width 0.01, 
are plotted in the lower frames of Fig.~\ref{v533her-ps} on the same scale. 
The figure (i)
confirms the larger amplitude of the superhump, (ii) reveals an
asymetrical shape of both waveforms with a lower rise and quicker drop of 
the superhump and vice versa for the orbital modulation, and (iii) shows that
the orbital waveform does not change neither in form nor in amplitude over
the time scale of $\approx$1~yr. 


The waveforms not being sinusoidal, overtones of the fundamental frequencies
are expected in the power spectra. Indeed, several peaks (much fainter than
the main signals) are seen at multiples of $F_{\rm orb}$ and $F_2$ and at
simple arithmetic combinations.
No other independent coherent periodicity 
could convincingly be detected up to the Nyquist frequency. In particular,
just as \citet{Leichty22} I see
no trace of the negative superhump at $P_{\rm nSH} = 0.142\,89$~d seen in 2004 by 
\citet{McQuillin12}. It therefore must have vanished in the meantime. 
Finally, I mention that similar to
the case of RZ~Gru, an enhancement of power between roughly 
35 -- 80~d$^{-1}$ indicates a broad range of QPOs in V533~Her.

\subsection{V795 Her: First view of a negative superhump}

V795~Her was discovered in the Palomar Green survey \citep{Green82} and
initially suspected to be an intermediate polar \citep{Zhang91, 
Thorstensen96, Shafter90}. However, it appears now to be
consensus that V795~Her is a SW~Sex type novalike system as first
proposed by \citet{Casares96}. With an orbital period of 0.108\,264\,8~(3)~d
determined spectroscopically by \citet{Shafter90}
it lies right within the period gap of cataclysmic variables. 
V795~Her shows photometric variations with a slightly varying period,
averaging 0.116\,19~(7)~d, somewhat longer than the spectroscopic 
period \citep[see, e.g.,][]{Kaluzny89, Patterson94, Papadaki06, Shafter90, 
Simon12}, establishing the system as a permanent superhumper. 
Superposed upon the superhumps are rapid flickering variations and
occasional QPOs with periods of 10 -- 20~min \citep{Rosen95, Patterson94,
Papadaki06}.

\begin{figure}
	\includegraphics[width=\columnwidth]{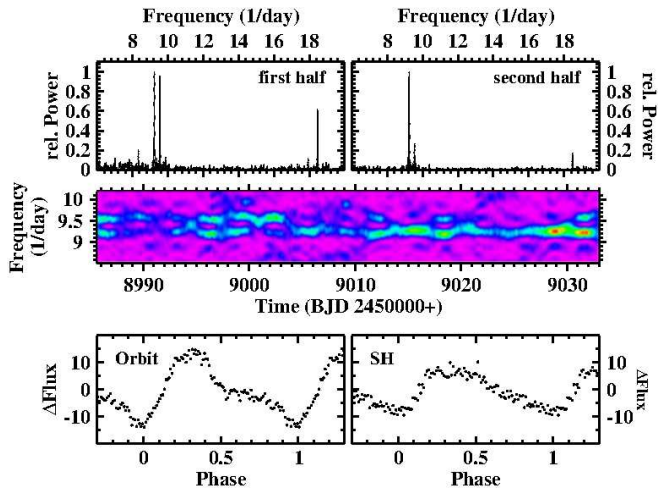}
      \caption[]{{\it Top:} Power spectrum of the first (left) and second
                half (right) of the light curve of V795~Her.
                {\it Middle:} Time resolved power spectrum of the entire
                light curve
                {\it Bottom:} Waveform of the orbital (left) and negative
                superhump variations (right) of V795~Her.}
\label{v795her-lc}
\end{figure}

The power spectrum of the TESS light curve is dominated by a signal at 
$F_{\rm orb} = 9.2362\, (9)$~d$^{-1}$ ($P_{\rm orb} = 0.108\,26\, (1)$~d),
which is the orbital frequency, and a second one at
a slightly higher frequency of
$F_{\rm nSH} = 9.547\, (1)$~d$^{-1}$ ($P_{\rm nSH} = 0.104\,74\, (1)$~d)
which I identify with a negative superhump. The first overtone of $F_{\rm orb}$
is also present, but not of $F_{\rm nSH}$. However, the relative strength of
the signals varies considerably over time. The upper frames of 
Fig.~\ref{v795her-lc} contain the power spectra of the first (left) and the
second half (right) of the entire light curve. While during the first half
both signals have similar strength, the orbital signal is much stronger than
the superhump signal during the second half. This is also obvious in the
time resolved power spectrum reproduced in the middle frame of the figure.

Not a trace of the positive superhump, expected at 8.61~d$^{-1}$ and seen in 
so many earlier observations, is present in the TESS data. On the other
hand, a negative superhump has never before been reported in this system.
The waveforms of the orbital and superhump variations are shown in the
lower frames of Fig.~\ref{v795her-lc}.

Finally, I mention an enhancement of power in the broad frequency range
of $\approx 65 - 125$~d$^{-1}$, consistent with the range of periods of
QPOs reported previously.

\subsection{HR Lyr: A 0.9~d modulation?}

HR~Lyr has erupted as a nova in 1919. The long term photometric
behaviour of the system was studied by \citet{Leibowitz95} on the base
of 5~yr of observations. They found a 64~d modulation. Observing HR~Lyr
for even 22~yr, \citet{Honeycutt14} could not confirm this period and
discuss possible reasons for this discrepancy. Instead, they find sections
of extended ascending or descending ramps in the brightness of HR~Lyr.
On nightly time scales, \citet{Leibowitz95} claim the presence of QPOs
around 0.1~h which they tentatively associate to an orbital 
period of about 2.4~h. However, it does not clearly reveal itself in
the light curves. The featureless spectrum of HR~Lyr 
\citep{Kraft64, Williams83} makes it almost impossible to measure the 
orbital period spectroscopically. Thus, it remains unknown.

\begin{figure}
	\includegraphics[width=\columnwidth]{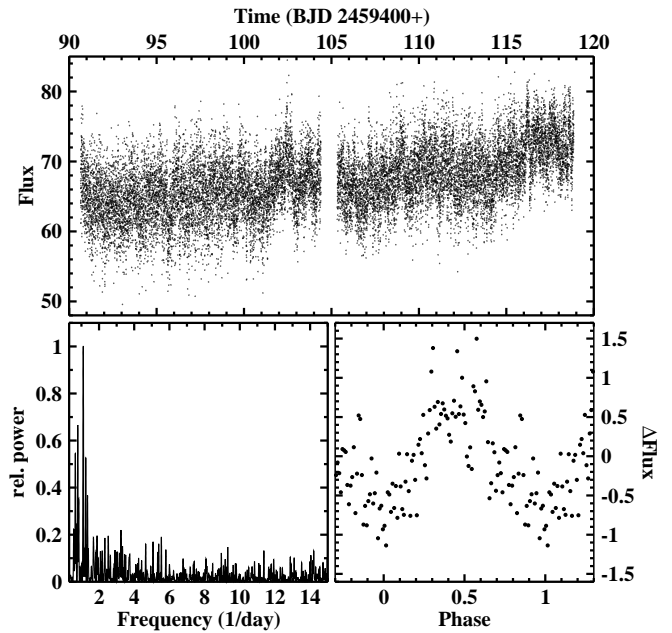}
      \caption[]{{\it Top:} Light curve of HR~Lyr.
                {\it Bottom left:} Power spectrum of the light curve after
                removing variations on time scales $>$4~d. 
                {\it Bottom right:} Light curve folded on the period
                0.907~d.}
\label{hrlyr}
\end{figure}

The TESS light curve of HR~Lyr is reproduced in
Fig.~\ref{hrlyr}. It exhibits a slight rise over the 28~d time base which
could be interpreted as a ramp as observed by \citet{Honeycutt14}. This notion
is confirmed by a check of the AAVSO long-term light curve of the
system. No periods are evident on a visual inspection.

The power spectrum (lower left frame of the figure) is almost featureless
up to very high frequencies. In particular, there is no indication for
the QPOs around $F=10$~d$^{-1}$ mentioned by \citet{Leibowitz95}.
Only at quite low frequencies a peak at
$F = 1.102\, (4) $~d$^{-1}$ ($P = 0.907\, (3)$~d) 
stands out moderately high above 
surrounding peaks. Considering this signal tentatively to represent a real
periodic variation of HR~Lyr, the data folded on this period yield the 
quite noisy low amplitude waveform shown in the lower right frame of 
Fig.~\ref{hrlyr}.

If real, could this period reflect the orbital motion? Probably not. At a
period of just short of 1~d the secondary star should contribute a significant
fraction of the system brightness and thus should readily be visible in the
spectrum of HR~Lyr. This is not the case not only when the system is on a
relatively high brightness level as during earlier spectroscopic 
observations cited above, but also not when it was at a temporarily 
much reduced brightness of $V \approx 17$ \citep{Munari16}. \citet{Leibowitz95}
speculate about the presence of a third body in HR~Lyr with an estimated
orbital period of around 1~d and consider worthwhile the search of a 
corresponding periodicity. Identifying the 0.9~d period suggested here with
the period predicted by them would pile speculation upon speculation. Thus,
I just mention this possibility here without claiming it to a have a sound
basis.

\subsection{MV Lyr: First observation of a strong negative superhump}

MV~Lyr is a well studied NL of the VY~Scl subtype. 
The orbital period has been determined spectroscopically by 
\citet{Voikhanskaya88}, \citet{Borisov92} and last by \citet{Skillman95},
but at 0.1329 (4)~d it remains of rather low precision.
MV~Lyr also exhibits a slightly longer photometric period
\citep{Borisov92, Skillman95} which may be interpreted as being due to a 
(positive) superhump. Periodic variations on longer times scales of
near 3.8~d \citep{Borisov92} and 3.6~(3)~d \citep{Skillman95} were also
reported but remain inconclusive because of the limited time base of the
respective observations.

Three TESS light curves of MV~Lyr are available.
Vastly more observations were obtained by the
Kepler satellite which monitored the star almost uninterruptedly in high
cadence mode for just short of 4 years. These data are used here to
complement the TESS data.

\begin{figure}
	\includegraphics[width=\columnwidth]{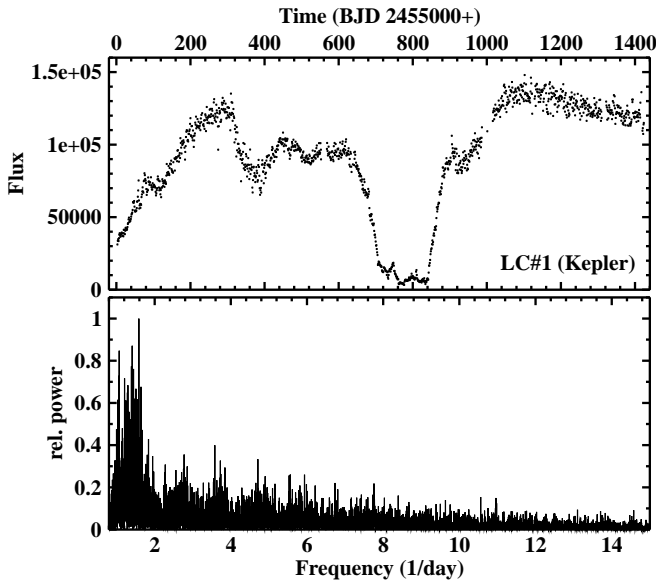}
      \caption[]{{\it Top:} Kepler light curve of MV~Lyr binned in 1-d
                 intervals.
                 {\it Bottom:} Power spectrum of the light curve.}
\label{mvlyr-kepler}
\end{figure}

Starting the discussion with the Kepler observations, the entire light
curve, binned in 1~d intervals for clarity, is shown in the upper frame
of Fig.~\ref{mvlyr-kepler}. For most of the time MV~Lyr remained in a high
state exhibiting gradual variations on the time scale of many days. For an
interval of about 200~d between May and October 2011 the high state was
interruped by a low state. Note that the considerable scatter of the
average daily flux is not due to noise but reflects real brightness
variations of MV~Lyr on the time scale of a day.

The power spectrum of the entire Kepler light curve is reproduced in the
lower frame of Fig.~\ref{mvlyr-kepler}. Is does not contain any outstanding
signal which would indicate a persistent periodicity. Investigating smaller
sections of the light curve, the power spectra sometimes contain marginally
significant peaks, but they are not repeated in other sections and cannot 
be associated to previously observed periodicities. Instead,
the power spectrum contains a wavelike structure with a dominant region of
many peaks centred on $\approx$1.4~d$^{-1}$, and other regions with decreasing
power at integer multiples of the main range. 

\begin{figure}
	\includegraphics[width=\columnwidth]{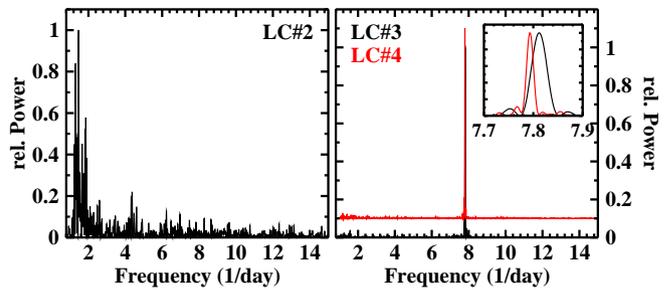}
      \caption[]{Power spectrum of TESS light curve LC\#2 (left) and light
                 curves LC\#3 and LC\#4 (right; the latter shifted vertically 
                 of clarity) of MV~Lyr. 
                 The insert shows a small frequency range around
                 the dominating peaks in order to emphasize their slightly
                 different frequencies.} 
\label{mvlyr-tess}
\end{figure}

The power spectrum of LC\#2 of the TESS observations (left frame of 
Fig.~\ref{mvlyr-tess}) is not unlike that of the Kepler data. The same,
however, cannot be said of LC\#3 and LC\#4, the power spectra of which
are totally different (right frame of the figure). They are dominated by a 
strong maximum at a very similar, but not identical frequency as is
better seen when drawn on the expanded scale in the insert of the figure.
In LC\#3 the maximum peaks at 
$F = 7.8120\, (8)$~d$^{-1}$ ($P = 0.128\,01\, (1)$~d),
in LC\#4 at 
$F = 7.7933\, (4)$~d$^{-1}$ ($P = 0.128\,315\, (6)$~d). 
The period is thus
slightly variable and about 3.5\% smaller than the orbital period, strongly 
suggesting that the signal is due to a negative superhump in MV~Lyr. In 
both light curves the waveform (not shown) is very nearly sinusoidal. 

The change in behaviour of MV~Lyr with no regular variations in LC\#1
and LC\#2 (before 2019, September) and the appearance of a very strong and 
never before seen negative superhump in LC\#3 and LC\#4 (after 2020, May)
is remarkable. Moreover, neither the Kepler nor the TESS data contain
any indication for the positive superhump observed by \citet{Borisov92}
and \citet{Skillman95}. But both suspect a long term periodicity in
their light curves. Is it a coincidence that the beat period between the 
orbital and the negative superhump period of MV Lyr is 3.63~d, very close 
the to the claimed long-term period suspected by them, but not detected in any 
of the present data?

\subsection{AQ Men: An improved orbital period}

AQ~Men was detected in the Edinburgh-Cape Blue Object 
Survey as EC~05114-7955 by \citet{Chen01}
and classified as a cataclysmic variable. They suspected a dwarf nova
nature but did not exclude the possibility of AQ~Men being a novalike
variable. \citet{Armstrong13}
prefer the latter classification based on the absence of 
observed outbursts even many years after discovery and the compatible 
spectrum. From radial velocity variations covering only 1.5 cycles 
\citet{Chen01} derived a period of 3.12~h. In extensive photometric 
observations \citet{Armstrong13} found a similar but more precise period 
of 0.141\,471 (8)~d and suspect the
presence of grazing eclipses. They also found a supraorbital period of
3.78~d from which they inferred the presence of a negative superhump and which
also causes a marginally significant peak in their power spectrum at
7.33~d$^{-1}$.

TESS observed AQ~Men many times. The data can be combined into
8 light curves, as detailed in Table~\ref{Table: obs-log}. The first four
of them have recently been analyzed in detail by \citet{Ilkiewics21}. I
found that the additional data only confirm their results without adding
significant new aspects. Therefore, I restrict myself here to 
taking advantage of the
total time base of the TESS data (1064~d), much longer than the 33~d time
base of the observations of \citet{Armstrong13},
to improve the orbital period. For this purpose I fold the
light curves (separating combined light curves LC\#4, LC\#5 and LC\#8 into
their original constituents) and determine representative epochs of the 
eclipse centres close to the middle of each light curve. This yields eleven 
pairs of cycle numbers and eclipse epochs
(Table~3). A linear least squares 
fit then results in the ephemeris:
\begin{equation}
T_{\rm min} = BJD\, 2458340.21172\, (5) + 0.141\,468\,35\, (1) \times E
\end{equation}
where the errors are the formal fit errors.

\subsection{V1193 Ori: Superhumps with variable waveforms}

V1193~Ori was discovered by \citet{Hamuy86} who noted its similarity to 
cataclysmic variables. This classification was confirmed by \citet{Bond87} 
who showed that the system is a UX~UMa-type novalike variable. A spectroscopic
period measured by \citet{Ringwald94} is consistent with the photometric 
period of 0.165\,001~(1)~d derived by \citet{Papadaki06}. However, while the 
latter authors did not see any other period in their data, \citet{Ak05}
claim the presence of three different periods which they interpret as 
orbital [0.1430~d, assuming the periods of \citet{Ringwald94} and 
\citet{Papadaki06} to be 1/day aliases], a negative superhump
(0.1362~d), and the beat between the other two periods (2.98~d).

\begin{figure}
	\includegraphics[width=\columnwidth]{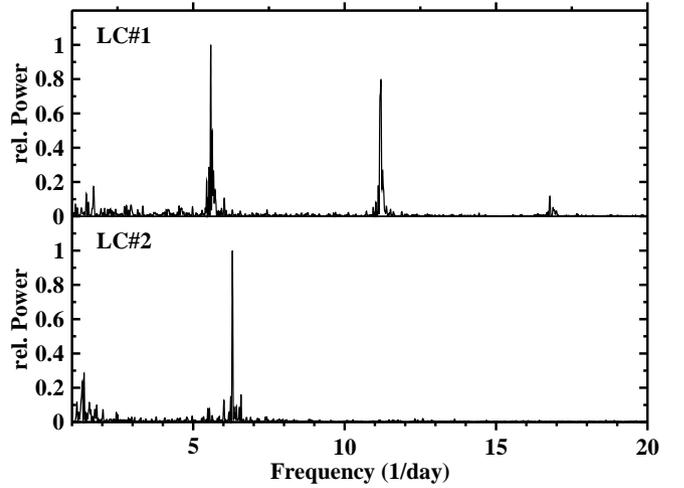}
      \caption[]{Power spectra of the two light curves of V1193~Ori.}
\label{v1193ori-ps}
\end{figure}

TESS obtained two light curves of V1193~Ori, separated by about 2~yr. 
They show only small variation on time scales longer than days. On shorter
time scales they contain clear variations which represent not only a challenge
to reconcile them with the previous observations cited above, but are also
drastically different during the two observing intervals. This is evident at
first glance, regarding the power spectra of the two light curves in
Fig.~\ref{v1193ori-ps}.

In both power spectra some signals are seen at low frequencies which, however,
do not have any relationship with other structures. I therefore consider them 
as being due to random variations on longer time scales. More importantly,
LC\#1 is dominated by a strong signal at 
$F_1 = 5.580\, (1)$~d$^{-1}$ ($P_1 = 0.179\,22\, (5)$~d). 
Clear overtones at integer \,multiples of $F_1$
are also present, but no other significant features. In contrast, LC\#2
only contains a single strong line at 
$F_2 = 6.296\, (1)$~d$^{-1}$ ($P_2 = 0.158\,83\, (4)$) with no overtones.  

\begin{figure}
	\includegraphics[width=\columnwidth]{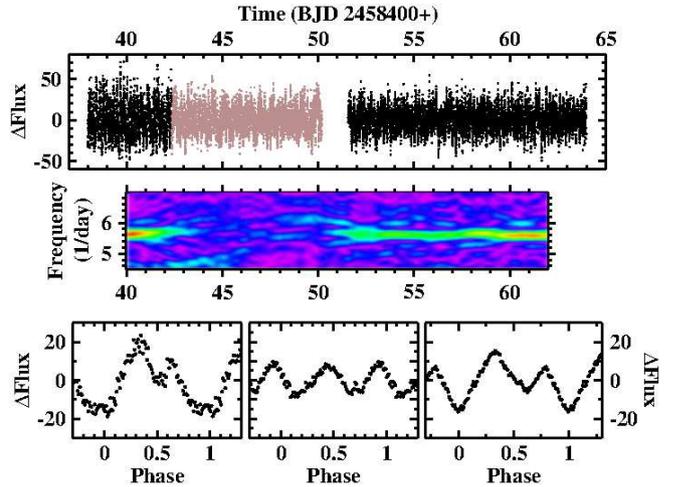}
      \caption[]{{\it Top:} LC\#1 of V1193~Ori (after subtraction of
                 variations on time scales $>$2~d) with a part exhibiting
                 different temporal characteristics (see text) plotted in a 
                 lighter shade. {\it Middle:} Time resolved power spectrum
                 in a narrow frequency range around the main signal at $F_1$.
                 {\it Bottom:} LC\#1 folded on $1/F_1$ and binned in phase
                 intervals of width 0.01: {\it (left:)} First part of light 
                 curve, {\it (middle:)} central (shaded) part, and 
                 {\it (right:)} final part.} 
\label{v1193ori-lc}
\end{figure}

A closer look at LC\#1 reveals another interesting feature. The light curve,
after subtraction of variations on time scales $>$2~d, is shown in the upper
frame of Fig.~\ref{v1193ori-lc}. A close inspection reveals even to the naked 
eye a regular pattern during the first $\approx$5~d. It reappears with a 
somewhat smaller
amplitude after the gap in the light curve. In between this pattern is not
readily apparent (shaded in the figure). In the time resolved power
spectrum in the middle frame of Fig.~\ref{v1193ori-lc} the otherwise strong
$F_1$ signal disappears during the highlighted part of the light curve. This
can be explained, regarding the waveform of the variations responsible for
$F_1$. The lower part of the figure contains the three parts of the light curve,
folded on $P_1$ and binned in phase intervals of width 0.01. The zero point
of phase is the same in all three frames, chosen to coincide with the minimum 
of the waveform of the second half of LC\#1 (after the gap). In the first and 
last part of LC\#1 the waveforms are highly structured but differ only slightly.
Apart from the difference in amplitude only a shift of the phase of the
second maximum is significant. The width of the principle minimum varies 
slightly, but its phase remains the same. During the central part of the light 
curve, however, the waveform changes drastically to very nearly sinusoidal with
a period $P_1/2$ and extrema at different phases than before and after.
The disappearance of $F_1$ in the time resolved power spectrum is then
explained by the doubling in frequency and the sinusoidal shape which
shifts all power to $2F_1$. This is confirmed by the time resolved power
spectrum around the first overtone of $F_1$ (not shown). 

In contrast, the waveform corresponding to the $F_2$ signal in LC\#2
is very nearly sinusoidal which explains the absence of overtones in
the power spectra. $F_2$ also appears in an independent data set. The
AAVSO archives contain a series of light curves of V1193~Ori, observed in
8 nights between 2006, Jan 8 -- 20. The power spectrum of the combined light 
curves (after subtracting the nightly average magnitude) has a strong peak 
at 
$6.291\, (4)$~d$^{-1}$, 
equal to $F_2$ within the error limits. The waveform of the corresponding 
variations is also quasi-sinusoidal.

How to interpret the various periods of V1193~Ori found here and in the
literature? Verifying the radial velocities listed in table 4 of
\citet{Ringwald94} I confirmed their spectroscopic orbital period and the
inconsistency of their data with either $P_1$ or $P_2$. Using the on-line
data provided by \citet{Papadaki06} I also confirmed the presence of only the
spectroscopic period in their measurements. In particular, the alternative
orbital period of 0.1430 proposed by \citet{Ak05} is incompatible with these
data. \citet{Ak05} based their analysis on a power 
spectrum with numerous peaks of similar height in the frequency range of
interest (see their Fig.~3). Their interpretation of structures corresponding
to periods of 0.1430~d and 0.1362~d as orbital and as a negative superhump
depends crucially on the assumption that the true orbital period is a 
1/d alias of the value derived by \citet{Ringwald94}. Moreover, they calculate
a probability of only 18\% that the alias period is correct. Taking into
account
 also
that the variations found by \citet{Papadaki06} must then be considered
as being due to a superhump which accidentally has a period exactly equal
to the originally proposed orbital period, leaves little evidence for the 
period of \citet{Ringwald94} to be wrong.

Coming back to the present data we then have a sequence 
$P_1 = 0.176\,223\,8\, {\rm d} > P_{\rm orb} = 0.165\,001\, {\rm d} >
P_2 = 0.158\,825\,2\, {\rm d}$ 
and
$\mid P_1 - P_{\rm orb} \mid / \mid P_2 - P_{\rm orb} \mid = 0.550$.
This suggests that $P_1$ is caused by a positive and $P_2$ by a negative
superhump in the light curve of V1193~Ori. But the 
variations seen by \citet{Ak05} remain unexplained.

\subsection{CP Pup: No ordinary old nova}
\label{CP Pup}

Nova Puppis 1942 (CP~Pup) was one of the brightest novae ever 
observed. Numerous studies have been dedicated to this system.
Yet its nature 
is still shrouded in mysteries. Not only are the spectroscopic and 
photometric periods different from each other, but various authors 
found discrepant values at different epochs \citep[see][and references 
therein for a more detailed discussion]{Bianchini12}. \citet{Mason13}
even raise doubts whether the
radial velocity variations reflect the binary period at all or whether
they might be caused by the rotation of a magnetic white dwarf 
in a system with a longer orbital period. A summary of published period
values is given in 
Table~5.

\begin{table}
\label{Table: CP Pup periods}
	\centering
	\caption{Spectroscopic and photometric periods (in days) of CP~Pup 
                 published in the literature. Errors of the last digits are
                 quoted in parentheses if provided in the original paper.}

\begin{tabular}{ll}
\hline
\multicolumn{2}{l}{Spectroscopic period} \\ [1ex]
0.06115       & \citet{Bianchini85} \\ 
0.061421 (25) & \citet{Duerbeck87}  \\
0.06141 (3)   & \citet{ODonoghue89} \\
0.06148 (5)   & \citet{ODonoghue89} \\
0.061375      & \citet{Bianchini90} \\
0.06129 (1)   & \citet{White93}     \\
0.06126454 (9)& \citet{Bianchini12} \\
0.0614466     & \citet{Mason13}     \\ [2ex]
\multicolumn{2}{l}{Photometric period} \\ [1ex] 
0.06196     & \citet{Warner85} \\ 
0.06198     & \citet{ODonoghue89} \\
0.06248     & \citet{ODonoghue89} \\
0.06154     & \citet{ODonoghue89} \\
0.0641      & \citet{Diaz91}     \\
0.06834 (7) & \citet{White93}    \\
0.061389 (5)& this paper         \\
\hline
\end{tabular}
\end{table}

\begin{figure}
	\includegraphics[width=\columnwidth]{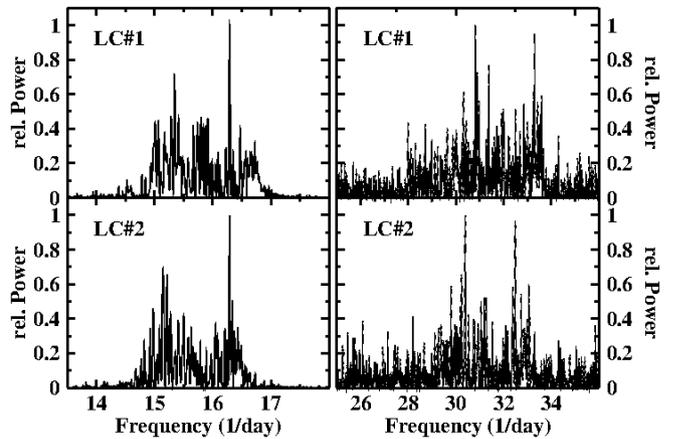}
      \caption[]{Power spectra of CP~Pup.}
\label{cppup-ps}
\end{figure}

Two TESS light curves of CP~Pup, separated by about 2~yr, are available. 
Some sections are evidently corrupted and/or contain short lived small 
scale outburst and were masked. The power spectra of both light curves are
similar but differ drastically from those of all other stars in this
study. They are dominated by a multitude of signals confined to the narrow 
frequency range between 14.3 and 17.0~d$^{-1}$ (left column of
Fig.~\ref{cppup-ps}), corresponding to periods between 0.059 and 0.070~d, 
i.e., encompassing the range of photometric periods reported in the past. 
More signals on a 1.5 orders of
magnitude lower power level occur at frequencies corresponding to the
first overtone of this range (right column of the figure). 
No other significant features were detected in the power spectra.
 
Only one individual peak, and then the strongest one, repeats itself in 
both light curves at exactly the same frequency within the error limits. In
LC\#2 its first overtone can also clearly be identified, but not in LC\#1.
Its average frequency is 
$F_1 = 16.290\, (1)$~d$^{-1}$ ($P_1 = 0.061\,389\, (5)$~d).
The reapparence of this signal and the
constancy of its frequency in two light curves separated by two years permits
to presume that it is a persistent periodic variation in CP~Pup. 

\begin{figure}
	\includegraphics[width=\columnwidth]{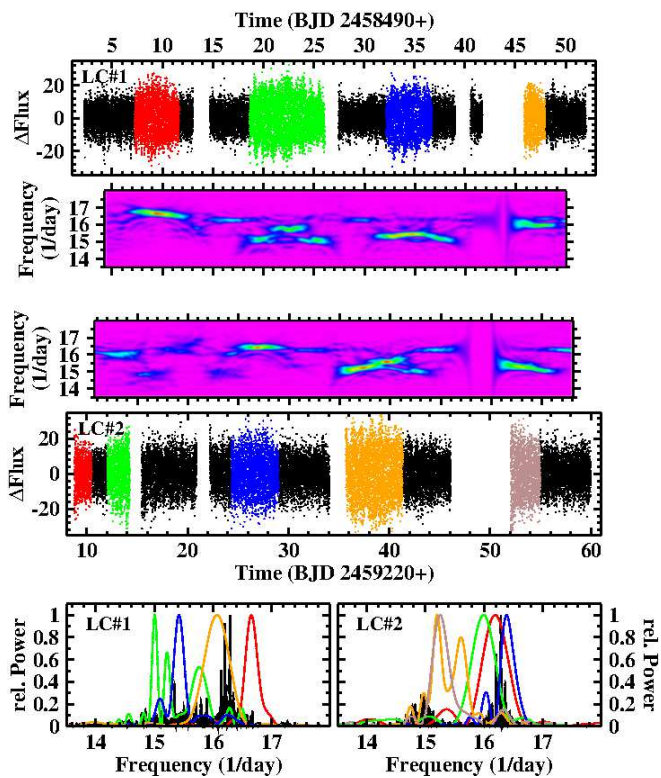}
      \caption[]{Time dependent behaviour of periodic variations in the
                 light curves of CP Pup. The second and third frames
                 contain the time dependent power spectra of light
                 curves LC\#1 and LC\#2. The light curves are shown
                 above and below, where quiescent and flaring states are
                 drawn in black and in different colours, respectively. The
                 power spectra of the different states are presented
                 in the lower frames, using the same colour code as in the
                 light curves.}
\label{cppup-lc}
\end{figure}

The odd apparence of the power spectrum becomes more transparent if its
time dependence is regarded. The second and third frames of 
Fig.~\ref{cppup-lc} contain the time resolved power spectra of the two
light curves in the relevant frequency range. A sliding window with a 
width of 4~d has been used to construct these spectra. Thus, 
structures within intervals of 4~d are not independent. The power spectra
are characterized
by a more or less continuous signal at $F_1$, often interrupted or supplemented
by other stronger and short-lived signals at neighbouring frequencies.
Such intervals are readily identified in the light curves which are plotted
above and below, respectively, the power spectra. Different states occur:
Most of the time CP~Pup remains in a state with a smaller amplitude of the
variations (black dots in the figure) which I will call quiescence. However,
this state is often interrupted by anomalous states of higher variability 
-- the flaring
states -- which are highlighted by different colours. These flaring states
coincide with the appearance of signals at frequencies other than $F_1$ in
the time resolved power spectra. In the lower frames of Fig.~\ref{cppup-lc}
the power spectra of only the quiescence (black) and the flaring states 
(using the same colour
code as used in the light curves) are shown. It is obvious that the $F_1$
signal is dominant in the quiescent state, while during flaring states
variations at other frequencies prevail. The waveform of the variations
is always very nearly sinusoidal.

The long and continuous TESS light curves of CP~Pup for the first time 
permit a more comprehensive characterization of its photometric variation 
than was possible with the much more limited terrestrial data sets of the
past. It appears clear that only one periodic signal at $P_1 = 0.016\,389$~d is
persistent, although it may be suppressed during flaring states when
variations at other periods become stronger. This period is close to 
published spectroscopic periods. It may thus well be the orbital period. 
But in view of the uncertainties in the interpretation of the spectroscopic
data, I am reluctant to associate $P_1$ definitely
to the orbital period of CP~Pup or the rotation period of the white
dwarf. Answering this and other deeper questions such as concerning the 
sudden appearance and disappearance of strong cyclic variations on a 
seemingly arbitrary period, albeit limited to a small and well defined 
period range, the physical origin of these always sinusoidal variations, or the 
reason for the period changes is beyond the scope of the present study.

The photometric behaviour of CP~Pup is clearly different from that of any
other CV. Although some of them sometimes exhibit bumps in their power
spectra attributable to QPOs, these cannot be compared to the narrow 
and well defined range of periods and the strength of the signals seen here. 
Thus, citing
the title of the paper of \citet{White93}: {\it CP Puppis: No ordinary old
nova}.

\section{Summary and outlook}

\begin{table*}
\label{Table: periods}	\centering
	\caption{Orbital, superhump and mutual beat periods (in days) 
                 identified in the TESS light 
                 curves of the target stars (literature values for the 
                 orbital periods are given whenever they are more accurate).}

\begin{tabular}{llllll}
\hline

Name & Type$^A$ & $P_{\rm orb}$ & $P_{\rm nSH}$ & $P_{\rm pSH}$ & $P_{\rm beat}$ \\
\hline
V795 Her  & SW       & 0.1082648~(3)$^f$ & 0.10474~(1) & -- & -- \\ 
DM Gem$^B$& N (1903) & 0.11570~(1)$^a$ & -- & 0.12423~(1) & -- \\ 
DM Gem$^B$& N (1903) & 0.12423~(1)$^a$ & 0.11570~(1) & -- & -- \\
MV Lyr    & VY       & 0.1329~(4)$^g$ & 0.12816~(1)$^C$ & -- & -- \\ 
TT Ari    & VY       & 0.13755040 (17)$^{c,D}$ & 0.132921~(2) & 0.15103~(2) & 1.1099~(6) \\ 
AQ Men    & UX       & 0.14146835~(1)$^a$ & -- & -- & -- \\ 
V533 Her  & N (1963) & 0.147374~(5)$^e$ & -- & 0.15645~(5) & -- \\ 
V704 And  & VY       & 0.15424~(3)$^{a,b}$ & 0.14772~(2) & -- & 3.49~(2) \\
V1193 Ori & UX       & 0.165001~(1)$^h$ & 0.15883~(4) & 0.179922~(5) & -- \\ 
TV Col    & IP       & 0.22860010~(2)$^a$ & 0.215995~(1) & -- & 3.895~(5) \\
AC Cnc    & UX       & 0.30047747~(4)$^d$ & 0.2824~(1) & -- & 4.65~(2) \\ 
RZ Gru    & UX       & 0.4175~(8)$^a$ & -- & 0.520(2) & -- \\
QU Car    & UX       & 0.4563~(2)$^a$ & -- & -- & -- \\ 
\hline
\multicolumn{6}{l}{$^A$N = nova (year or eruption); UX = UX~UMa; VY = VY~Scl; 
                       SW = SW~Sex;}\\
\multicolumn{6}{l}{\phantom{$^A$}IP = intermediate polar} \\
\multicolumn{6}{l}{$^B$Two alternative interpretations} \\
\multicolumn{6}{l}{$^C$Average from two light curves} \\
\multicolumn{6}{l}{$^D$orbital period not detected in TESS data} \\
\multicolumn{6}{l}{$^a$this work; $^b$\citet{Weil18}; $^c$\citet{Wu02};
$^d$\citet{Thoroughgood04};} \\
\multicolumn{6}{l}{$^e$\citet{Thorstensen00}; $^f$\citet{Shafter90};   
$^g$\citet{Skillman95}; } \\
\multicolumn{6}{l}{$^h$\citet{Papadaki06}  } \\
\end{tabular}
\end{table*}

The value of uninterrupted long duration and high time resolution
photometric observations from space performed in particular by the
Kepler and TESS missions not only but in particular for 
CV research has been demonstrated many times in the past. The current
study can only confirm this obvious statement. In all of the 15 objects
investigated here some new and sometimes surprising aspects have been 
found (although admittedly the current sample is biased due to the
selection critera mentioned in Sect.~\ref{Introduction}).
An abundance of superhumps, positive as well as negative, was detected,
many of which have never been seen before. Some of them exhibit an
interesting behaviour concerning their temporal development or their
waveforms. Their periods are summarized 
together with some newly derived or improved orbital periods in
Table~6, ordered according to increasing orbital period.
There is no preference for superhumps to occur, or not to occur, among
specific CV types or in a limited period range. It appears that
negative superhumps are more common than positive ones, but this is not
a statistically sound statement. Thus, the limited sample does not permit 
to draw clear and meaningful conclusions. A broader basis for a systematic
accessment of superhumps in CVs other than outbursting dwarf novae is
required. This is expected to be achieved by the extension of the current
study, currently in progress as mentioned in the introduction, to more 
systems known to have exhibited superhumps in the past. A more thorough
discussion, also taking into account the large amount of relevant information
scattered in the literature, is therefore postponed to the follow-up study.

Even so, the availability of larger and better datasets already
demonstrates that the occurrence of superhumps in novalike variables and
old nova is not the exception as may have been expected in the past, but is
a rather common feature in the light curves of these systems.
Worthwhile mentioning is also the first
detection in optical light of the WD spin period in TV~Col in form of 
its orbital side band, and the curious time dependent frequency 
behaviour of CP~Pup which explains the multitude of apparently contradicting
periods reported in the past. Both results could not have been obtained
easily by terrestrial observations.

This is an exploratory study. It has the purpose to identify interesting
variations in the target stars and to document them. An in-depth investigation
of their origin is left to subsequent work. There are many more CVs with TESS
light curves which deserve to be scrutinized. The work is going on.

\section*{Acknowledgements}

This paper is based on data collected by the TESS and Kepler missions and 
obtained from the MAST data archive at the Space Telescope Science Institute 
(STScI). Funding for the missions is provided by the NASA Explorer Program
and the NASA Science Mission Directorate for TESS and Kepler, respectively. 
STScI is operated by the Association of Universities for Research in 
Astronomy, Inc., under NASA contract NAS 5-26555. Supportive data were
obtained from the data arquives operated by the American Association of
Variable Star Observers. I am grateful to Alexandre de Oliveira for his
advice on the handling of TESS data.

\section*{Data availability}

All data used in the present study are publically available at the
Barbara A.\ Mikulski Archive for Space Telescopes 
(MAST):\\ https://mast.stsci.edu/portal/Mashub/clients/MAST/Portal.html
and at the AAVSO web site: https://www.aavso.org.






\bsp	
\label{lastpage}
\end{document}